\documentclass%
{amsart}
\usepackage[foot]{amsaddr}

\usepackage[hmargin=30mm, top=25mm, bottom=30mm]{geometry}
\usepackage[letterspace=150]{microtype}
\usepackage[onehalfspacing]{setspace}

\usepackage[pagewise]{lineno}

\usepackage[english]{babel}
\usepackage{lmodern}
\usepackage[OT1]{fontenc}

\usepackage{graphicx}
\usepackage{adjustbox}
\usepackage{subcaption}
\usepackage{wrapfig}

\usepackage{pgfplots}
    \pgfplotsset{compat=newest}
\usepackage{tikz}
    \usetikzlibrary{arrows.meta, %
        bending, %
        chains, %
        decorations.pathreplacing, %
        fit, %
        positioning, %
        quotes, %
        scopes, %
        shapes}

\usepackage[compat=0.6]{yquant}
    \yquantset{register/default name=$\reg_{\idx}$}
    \yquantset{register/minimum height=10pt}
    \yquantset{register/separation=2pt}
    \yquantset{operator/separation=10pt}
    \yquantset{every circuit/.style={line width=0.5pt}}
    \yquantset{every operator/.style={scale=1.0, font=\normalsize}}
    \yquantset{every control/.style={shape=yquant-circle, anchor=center, radius=.65mm, draw}}
    \yquantset{operators/every not/.style={shape=yquant-oplus, radius=1.6mm, draw}}
    \yquantset{operators/every box/.style={%
        shape=yquant-rectangle, draw, align=center, inner timesep=1.2mm, time radius=2.8mm, space radius=2.8mm}}
    \yquantset{every label/.append style={font=\normalsize}}

\usepackage{xcolor}

\definecolor{CBblue}{rgb}{0.2667, 0.4667, 0.6667}
\definecolor{CBcyan}{rgb}{0.4, 0.8, 0.9333}
\definecolor{CBgreen}{rgb}{0.1333, 0.5333, 0.2}
\definecolor{CByellow}{rgb}{0.8, 0.7333, 0.2667}
\definecolor{CBred}{rgb}{0.9333, 0.4, 0.4667}
\definecolor{CBpurple}{rgb}{0.6667, 0.2, 0.4667}
\definecolor{CBgrey}{rgb}{0.7333, 0.7333, 0.7333}

\definecolor{CBpaleblue}{rgb}{0.7333, 0.8, 0.9333}
\definecolor{CBpalecyan}{rgb}{0.8, 0.9333, 1.0}
\definecolor{CBpalegreen}{rgb}{0.8, 0.8667, 0.6667}
\definecolor{CBpaleyellow}{rgb}{0.9333, 0.9333, 0.7333}
\definecolor{CBpalered}{rgb}{1.0, 0.8, 0.8}
\definecolor{CBpalegrey}{rgb}{0.8667, 0.8667, 0.8667}

\definecolor{CBdarkblue}{rgb}{0.1333, 0.1333, 0.3333}
\definecolor{CBdarkcyan}{rgb}{0.1333, 0.3333, 0.3333}
\definecolor{CBdarkgreen}{rgb}{0.1333, 0.3333, 0.1333}
\definecolor{CBdarkyellow}{rgb}{0.4, 0.4, 0.2}
\definecolor{CBdarkred}{rgb}{0.4, 0.2, 0.2}
\definecolor{CBdarkgrey}{rgb}{0.3333, 0.3333, 0.3333}

\usepackage{braket}
\usepackage{bm}
\usepackage{amsmath}
\usepackage{amsfonts}
\usepackage{amssymb}
\usepackage{amsthm}
\usepackage{mathtools}
\usepackage{braket}

\usepackage{datetime2}
\usepackage{float}
\usepackage{ragged2e}
\usepackage{csquotes}
\usepackage{placeins}
\usepackage{booktabs}
	
\usepackage{caption}

\usepackage[style=numeric-comp,giveninits=true]{biblatex}
\usepackage{hyperref}
\hypersetup{ 
    colorlinks=true,
    urlcolor=blue
    }

\newcommand{\dagg}[1]{{#1}^{\dagger}}
\DeclareMathOperator{\Tr}{Tr}

\newcommand{\cal}[1]{\mathcal{#1}}
\newcommand{\vu}{\mathbf{u}} 
\newcommand{\bigO}{\cal{O}} 
\newcommand{\reyn}{\operatorname{\mathit{R\kern-.04em e}}} 

\DeclareMathOperator*{\x}{X}
\DeclareMathOperator*{\z}{Z}
\DeclareMathOperator*{\cx}{CX}

\DeclareMathOperator*{\hgate}{H}

\newcommand{\code}[1]{\texttt{#1}}
\newcommand{\quotes}[1]{``#1''}

\title[Quantum algorithm for the LBM]{Quantum algorithm for the lattice Boltzmann method with applications on real quantum devices}

\newcommand{\withquanscient}{$^{\ast}$}
\newcommand{\withlut}{$^{\dagger}$}
\newcommand{\withhaiqu}{$^{\ddagger}$}
\newcommand{\withatip}{$^{\S}$}
\newcommand{\withamu}{$^{\P}$}
\newcommand{\withtecsci}{$^{\daleth}$}

\author[A. Bastida-Zamora]{
    Antonio Bastida-Zamora\withquanscient}
\author[Lj. Budinski]{
    Ljubomir Budinski\withquanscient\withtecsci}
\author[O. Kerppo]{
    Oskari Kerppo\withquanscient}
\author[V. Lahtinen]{
    Valtteri Lahtinen\withquanscient\withlut}
\email{valtteri.lahtinen@quanscient.com}
\author[O. Niemimäki]{
    Ossi Niemimäki\withquanscient}
\email{ossi.niemimaki@quanscient.com}
\author[W. Steadman]{
    William Steadman\withquanscient}
\author[R. Zamora-Zamora]{
    Roberto Zamora-Zamora\withquanscient}

\author[P. Sagaut]{
    Pierre Sagaut\withamu}

\author[V. Bohun]{Vladyslav Bohun\withhaiqu}
\author[M. Koch-Janusz]{Maciej Koch-Janusz\withhaiqu}
\author[I. Lukin]{Illia Lukin\withhaiqu\withatip}

\address{\withquanscient{}Quanscient Oy, Peltokatu 34, 33100 Tampere, Finland}
\address{\withlut{}School of Engineering Science, Lappeenranta–Lahti University of Technology, P.O. Box 20, 53851 Lappeenranta, Finland}
\address{\withhaiqu{}Haiqu, Inc., 95 Third Street, San Francisco, CA 94103, USA}
\address{\withatip{}Akhiezer Institute for Theoretical Physics, NSC KIPT, Akademichna 1, 61108 Kharkiv, Ukraine}
\address{\withamu{}Aix Marseille Univ, CNRS, Centrale Med, M2P2 UMR 7360, Marseille, France}
\address{\withtecsci{}University of Novi Sad, Faculty of Technical Sciences, Trg Dositeja Obradovica 6, Novi Sad, Serbia}

\date{\today}

\begin{document}

\begin{abstract}
We introduce a novel quantum algorithm for the lattice Boltzmann method (LBM) based on the one-step simplified LBM. The structure of the algorithm allows for more flexibility in modelling different physics in contrast to earlier quantum algorithms for the LBM, while retaining computational efficiency in terms of the gate and qubit complexity. The new algorithm has potential for full end-to-end quantum utility especially for linear problems. We discuss the implementation of examples in linear acoustics, as well as a nonlinear Navier-Stokes problem that was solved on an IBM QPU in a hybrid simulation loop.
\end{abstract}

\maketitle
\markboth{BASTIDA-ZAMORA ET AL.}{QUANTUM ALGORITHM FOR THE LBM}

\section{Introduction}%
\label{sec:cfd}

Computational fluid dynamics (CFD) has become a ubiquitous tool in engineering of daily use in almost all industrial fields, but also in life sciences, Earth sciences and astrophysics, with some exotic extensions to the simulation of many particles systems such as human crowds and diffusion of innovation.
Since the seminal works carried out at Los Alamos during the Manhattan project~\cite{vonneumann1944proposal,bethe1944,richtmyer1949modern,vonneumann1950method}, a vast number of methods have been proposed to discretize and solve the continuous governing equations (usually the Euler and Navier-Stokes equations), some of the most popular examples being finite difference, finite volume, finite element, and spectral methods~\cite{hirsch1990numerical1,hirsch1990numerical2,peyret2002spectral,sengupta2004fundamentals,moser2022numerical}. 
The development of the related numerical algorithms was mainly driven by efficiency, which is a mix of robustness, accuracy, and the ability to make the most of the properties of processing units.
This is well illustrated by the rise of time-marching explicit methods with very small time steps on computational grids with a huge number of grid points (or cells), which are very well suited to massively parallel computers based on processors with small memory and restricted computing power.
For the sake of comparison, implicit methods for steady-state computations were best suited for computers based on a single or very few very powerful processors, like the CRAY machines during the 90s. 
With the rise of computers integrating up to tens or hundreds of millions of GPUs or heterogeneous many-core processors, a new step in frontier computing has been reached~\cite{wilfong2025simulating}. This opens the way to increasingly realistic simulations of multiphysics, full-scale, and fully detailed systems in the presence of turbulence.

The search for methods well suited to massively parallel computers made the lattice Boltzmann method (LBM)~\cite{succi2001lattice,kruger2016lattice} popular, since its mathematical structure supports efficient parallelization. 
Fundamentally, the LBM appears as a set of coupled linear advection equations supplemented by local non-linear relaxation terms.
These equations are usually solved using a very compact spatial stencil along with an explicit two-step time integration method originating from a Strang splitting between advection and relaxation operators. These two key elements minimize the data exchange between processors, and make it easier to obtain linear speedup even when tens or hundreds of millions of processors are used.
This is illustrated by some landmark achievements, for example  \cite{xu2025towards} demonstrated simulations of urban physics with up to $2 \times 10^{12}$ cells using $155 \times 10^6$ heterogeneous many-core CPUs.

Another very interesting feature of LBMs is their versatility.
While they were first proposed as an empirical improvement of lattice gas automata (a type of cellular automata designed to mimic hydrodynamic phenomena) \cite{mcnamara1988use,higuera1989Boltzmann,higuera1989lattice,qian1992lattice}, it was later proved that at least some of them can be rigorously derived as discretizations of the classical Boltzmann equation \cite{he1997theory,abe1997derivation,shan2006kinetic}. The last but very important step was the demonstration that they can be seen as a smart and efficient way to solve partial differential equations (PDEs) in general \cite{chai2018lattice,bi2025lattice,yang2022unified}, opening the way to their use in solid mechanics \cite{obrien2012lattice,murthy2018lattice,escande2020lattice,boolakee2025lattice,muller2024improvement}, in combustion \cite{yamamoto2002simulation,boivin2021benchmarking,taileb2022lattice,wissocq2023hybrid,hosseini2023low}, in heat transfer \cite{jiaung2001lattice,wei2023unified}, in electrodynamics \cite{mendoza2010three,hanasoge2011lattice,hauser2017stable}, and in quantum physics \cite{succi1993lattice,zhong2006lattice,lapitski2011convergence,inui2021coupled,wang2017numerical}, but also, in more exotic topics such as pedestrian and traffic flows \cite{meng2008modeling,meng2008lattice,shi2016revised}, crowd dynamics \cite{xue2017abnormal}, and epidemics \cite{de2020modeling}.

Despite the massive simulation capabilities provided by modern-day high-performance computing and methods such as LBM, many problems in CFD remain intractable or painfully resource-intensive. High Reynolds numbers and turbulence, intricate geometries, multiphase flows and multiphysics coupling can pose insurmountable limitations for carrying out CFD even with the best-in-class computer clusters. For instance, turbulence has a $\bigO(\reyn^3)$ space-time dependence on the Reynolds number $\reyn$, and full airplane simulations feature Reynolds numbers in the scale of $10^8$, requiring roughly $10^{24}$ floating-point operations for accurate solution~\cite{succi2023quantum}. As Moore's law reaches its physical limits in silicon chips due to atomic-scale constraints, quantum computing emerges as a promising, and possibly the one and only alternative, for computationally intensive fields like CFD \cite{bohun2025quantum,xiu2019time,rupp2010economic,theis2017end,ezratty2023there}.

Quantum computing can avoid the resource bottlenecks associated with ordinary computers by utilizing quantum-native features such as entanglement, interference, and superposition. An efficient quantum algorithm could provide an exponential number of degrees of freedom with respect to a linear qubit increase, while requiring far fewer computational operations than the corresponding classical method.
Such speedups could lead to significantly more complex simulations to be realized with quantum computers. Moreover, the inherent compactness and lower power requirements of some quantum computing architectures can help make the growing computation demands more feasible and environmentally friendly.

Numerous quantum algorithms for solving PDEs, and for CFD in particular, have been proposed. Many of these are based on, or are variants of, quantum linear system solvers, such as the HHL~\cite{hhl} or the variational quantum linear solver~\cite{cerezo2021}. Others are based on emergent nonlinear dynamics of interacting particles, such as lattice gas automata \cite{georgescu2025,Fonio2025QuantumLGCA,Fonio2026QuantumLGABurgers,BastidaZamora2025EfficientQLGA,Zamora2025FloatingPointQLGA}. HHL is based on quantum phase estimation, a subroutine where the computational complexity scales as $\bigO(1/\epsilon)$ with $\epsilon$ the precision to achieve. Furthermore, the method scales as $\bigO(\kappa^2)$ with $\kappa$ the condition number of the operator (ratio between largest and smallest eigenvalue). For this reason, HHL is only efficient for operators for which the condition number is polylogarithmic with respect to the number of grid points. 
In the context of discretized PDEs, errors incurred by the quantum phase estimation need to be lower than the mesh discretization errors, while lower errors require a higher number of qubits. The condition number increases with higher grid resolutions, creating serious problems in scaling the simulation~\cite{jin2022Time}. To solve time-stepping problems typical in CFD, one would also need to embed the time-dimension into the linear system by encoding all the steps required, for example by using a finite difference formulation with time-parametrized superposition states~\cite{li2025potential}. 
In variational quantum algorithms, on the other hand, iteration between quantum and classical computations creates a bottleneck for large and long simulations where the state preparation and measurements have an exponential complexity with the number of grid points. Furthermore, variational quantum algorithms for complex problems can lead to very deep quantum circuits due to barren plateaus~\cite{larocca2025Barren}. 

Quantum algorithm for the lattice Boltzmann method offers an alternative to the linear system solvers for CFD. In general, the LBM circumvents the direct solution of a linear system by formulating the field evolution as the propagation and collision of mesoscopic probability distributions of fictive particles, from which the the PDE solution can be recovered. This structure makes the LBM particularly appealing for a quantum algorithm implementation: simplified collision processes can be localized to be independent of lattice sites, while the propagation can be captured as a quantum walk~\cite{succi2015walk}. The heavy memory requirement that tends to be the bottleneck for the classical LBM is no real concern in quantum implementations. Thus various quantum algorithms for the LBM have been suggested, to varying degrees of generality~\cite{budinski2021, budinski2022,
georgescu2025, schalkers2024momentum, xiao2025quantum, wawrzyniak2025dynamic, sanavio2024carleman, sanavio2024three, sanavio2025carleman}.

%

In this paper, we propose a new quantum algorithm for the lattice Boltzman method. This algorithm is based on the one-step simplified LBM (OSSLBM), a variant of the LBM with a two-level propagation process~\cite{qin2022}. As the propagation is based on the efficient quantum basis shift subroutine~\cite{shift-preprint}, the increase in circuit complexity is negligible when balanced with the increased flexibility of the model in capturing different physics.
Propagation is also the only part of the algorithm where we require distribution functions; otherwise, the operators deal directly with the macroscopic lattice fields. This simplifies the collision and the boundary operators in particular. While the OSSLBM does not have collision in the sense of the standard LBM, there is nevertheless a collision-like operator mapping the lattice fields to distribution functions tied to local discrete velocities. In particular, this operator retains locality.

We will discuss in detail how a quantum algorithm for the OSSLBM represents a true opportunity for end-to-end quantum advantage. Moreover, the ability to implement different physical processes yields a wide variety of applications, and we highlight a few examples. 

In Section~\ref{sec:lbm}, we briefly introduce LBM, focusing in particular on the one-step simplified model. Section~\ref{sec:quantumlbm} describes the subroutines in the quantum algorithm in general terms. In Section~\ref{sec:analysis}, we discuss the scope of the method and carry out a brief complexity analysis in terms of the gate and qubit resources required, as well as discuss the full simulation over several time-steps. We also show some numerical examples including linear acoustics models in Section~\ref{sec:example_acoustics}, and a nonlinear Navier-Stokes simulation on an IBM quantum computer in Section~\ref{sec:example_airfoil}, before taking a brief look into the future of the quantum LBM in Section~\ref{sec:future}.
\section{Lattice Boltzmann method and previous work on quantum LBM}%
\label{sec:lbm}

This section is devoted to a brief reminder about fundamentals of lattice Boltzmann methods, the emphasis being put on the derivation of the one-step simplified method starting from the collide-and-stream process.
In particular, we sketch the move from generic collision integral to approximate collision operator by Bhatnagar–Gross–Krook (BGK)~\cite{BGK}, then to a simplification of the distribution evolution, and finally to a distillation of the underlying two-step process to a single step involving more variables.

\subsection{Lattice Boltzmann with the BGK collision}

Since the OSSLBM is based on rewriting the BGK collision model, a brief reminder will be provided here for the sake of clarity and self-consistency. The BGK is an important simplification allowing to model a wide selection of different physics, and -- importantly for numerical implementations -- reducing the collision to a local operator. Note, however, that the BGK is not the answer for all models, and more general operators may be necessary~\cite[Ch.~10]{kruger2016lattice}.

The construction begins with a lattice-based discretization.
We first define a set of discrete velocities $\bm{c}_{\alpha}$ with associated distribution functions $f_{\alpha}$, where $\alpha \in \{0, \dots, m-1\}$. This, together with the BGK model, yields the discrete velocity Boltzmann equation:
\begin{equation}
    \frac{\partial f_{\alpha}}{\partial t} + \bm{c}_{\alpha} \cdot \nabla f_{\alpha} = - \frac{1}{\tau} \left(f_{\alpha}(\bm{x},t) - f^{eq}_{\alpha}(\bm{x},t)\right) ,
    \label{eq:discrete-boltzmann}
\end{equation}
which, on a D$n$Q$m$ lattice, can be written as:
\begin{equation}
    f_{\alpha}(\bm{x}+\bm{c}_{\alpha}\Delta t, t+\Delta t) = f_{\alpha}(\bm{x},t) + \frac{\Delta t}{\tau}\left(f_{\alpha}(\bm{x},t) - f^{eq}_{\alpha}(\bm{x},t)\right) .
    \label{eq:lbm-bgk}
\end{equation}
The term on the right-hand side is the BGK linear approximation of the general collision integral, using the equilibrium distribution functions $f^{eq}_{\alpha}(\bm{x}, t)$, the relaxation time $\tau$ and the integration time step $\Delta t$. Note that the expression for collision is purely local.

We can then implement the collide-and-stream process as follows
\begin{align}
    f_{\alpha}^{coll}(\bm{x},t) &= f_{\alpha}^{eq}(\bm{x},t) + \left( 1 - \frac{\Delta t}{\tau}\right)f_{\alpha}^{neq}(\bm{x},t) 
    && \text{(collision)}
    \label{eq:collision}
    \\
    f_{\alpha}(\bm{x}+\bm{c}_{\alpha}\Delta t,t+\Delta t) &= f_{\alpha}^{coll}(\bm{x},t) && \text{(streaming)}
    \label{eq:streaming}
\end{align}
where $f^{neq}_{\alpha}(\bm{x}, t)$ is the non-equilibrium part of the distribution function $f_{\alpha}$. Exact expressions of these terms are dependent on the collision model.

The associated fluid viscosity is $\nu = c^2_s\Delta t \left(\tau - 1/2\right)$, where $c_s = c_s^*\Delta x / \Delta t$ is the speed of sound, with $\Delta x$ the mesh size and $c_s^*$ is the built-in lattice speed. The macroscopic quantities $V_0$ and $\bm{V}_1$ (e.g. density $\rho$ and momentum $\rho \bm{u}$ in the classical hydrodynamic case, see below) are reconstructed in the following way:
\begin{equation}
\begin{aligned}
    V_0 &= \sum_{\alpha} f_{\alpha} = \sum_{\alpha} f_{\alpha}^{eq}, \\
    \bm{V}_1 &= \sum_{\alpha} \bm{c}_{\alpha} f_{\alpha}= \sum_{\alpha} \bm{c}_{\alpha} f_{\alpha}^{eq} .
    \label{eqn_5}
\end{aligned}
\end{equation}
The main characteristics of the usual LBM lattices (see Figure~\ref{fig:D2Q9 and D2Q17 lattice}, and \cite[Ch.~5]{wolfgladrow2004},\cite[Ch.~3.4]{kruger2016lattice} for more detailed discussion) are given in Table~\ref{tab:lattice-features}.
\begin{table}[!ht]
    \centering
    \caption{Main features of the D2Q9 and D3Q27 lattices for lattice Boltzmann methods. The characteristic speed of the lattice is $c_s=1/\sqrt{3}$ in both cases.}
    \label{tab:lattice-features}
    \begin{tabular}{c|c|c|c}
       Lattice  & $\bm{c}_{\alpha}$ & Number &  Weight $W_\alpha$ \\ \hline
       D2Q9 & $(0,0)$ & 1 &  4/9 \\
            & $(\pm 1,0), (0,\pm 1)$ & 4 & 1/9 \\
            & $(\pm 1,\pm 1)$ & 4 & 1/36 \\ \hline
        D3Q27 & $(0,0,0)$ & 1 &  8/27 \\
            & $(\pm 1,0,0), (0,\pm 1, 0), (0,0, \pm 1)$ & 6 & 2/27 \\
            & $(\pm 1,\pm 1,0), (0,\pm 1, \pm 1), (\pm 1,0, \pm 1)$ & 12 & 1/54 \\
            & $(\pm 1, \pm 1, \pm 1)$ & 8 &  1/216 \\ \hline
    \end{tabular}
\end{table}
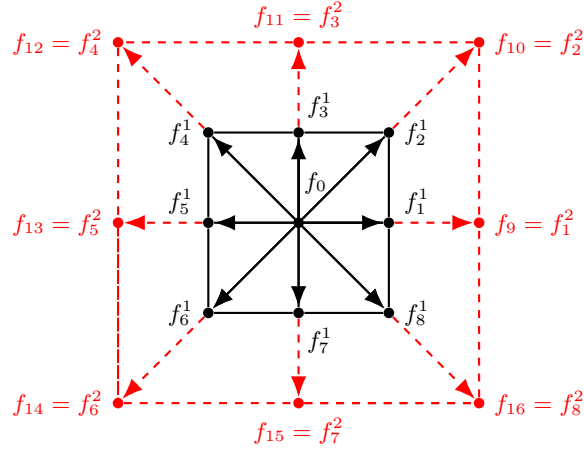
\begin{figure}[htb]
	\centering
	\begin{tikzpicture}[scale=0.4]
    \tikzset{
        lattice node/.style={fill=black, circle, minimum size=4pt, inner sep=0pt},
        lattice node outer/.style={fill=red, circle, minimum size=4pt, inner sep=0pt},
        node label/.style={font=\small},
        node label outer/.style={font=\small, color=red},
        lattice line/.style={black, thick},
        lattice line outer/.style={red, thick, dashed},
        arrow line/.style={black, thick, -{Latex[scale=1.2]}},
        arrow line outer/.style={red, thick, dashed, -{Latex[scale=1.2]}}
    }
   
    \node[lattice node] (f0) at (0,0) {}; 
    \node[node label, anchor=south, xshift=8pt, yshift=6pt] at (f0.north west) {$f_0$};
    
    \node[lattice node] (f2) at (3,3) {}; 
    \node[node label, anchor=west] at (f2.east) {$f_2^1$};
    
    \node[lattice node] (f1) at (3,0) {}; 
    \node[node label, anchor=west, yshift=8pt] at (f1.east) {$f_1^1$};
    
    \node[lattice node] (f3) at (0,3) {}; 
    \node[node label, anchor=south, xshift=8pt] at (f3.north) {$f_3^1$};
    
    \node[lattice node] (f4) at (-3,3) {}; 
    \node[node label, anchor=east] at (f4.west) {$f_4^1$};
    
    \node[lattice node] (f5) at (-3,0) {}; 
    \node[node label, anchor=east, yshift=8pt] at (f5.west) {$f_5^1$};
    
    \node[lattice node] (f6) at (-3,-3) {}; 
    \node[node label, anchor=east] at (f6.west) {$f_6^1$};
    
    \node[lattice node] (f7) at (0,-3) {}; 
    \node[node label, anchor=north, xshift=8pt] at (f7.south) {$f_7^1$};
    
    \node[lattice node] (f8) at (3,-3) {}; 
    \node[node label, anchor=west] at (f8.east) {$f_8^1$};
    
    \node[lattice node outer] (f10) at (6,6) {}; 
    \node[node label outer, anchor=west] at (f10.east) {$f_{10} = f_2^2$};
    
    \node[lattice node outer] (f9) at (6,0) {}; 
    \node[node label outer, anchor=west] at (f9.east) {$f_9 = f_1^2$};
    
    \node[lattice node outer] (f11) at (0,6) {}; 
    \node[node label outer, anchor=south] at (f11.north) {$f_{11} = f_3^2$};
    
    \node[lattice node outer] (f12) at (-6,6) {}; 
    \node[node label outer, anchor=east] at (f12.west) {$f_{12} = f_4^2$};
    
    \node[lattice node outer] (f13) at (-6,0) {}; 
    \node[node label outer, anchor=east] at (f13.west) {$f_{13} = f_5^2$};
    
    \node[lattice node outer] (f14) at (-6,-6) {}; 
    \node[node label outer, anchor=east] at (f14.west) {$f_{14} = f_6^2$};
    
    \node[lattice node outer] (f15) at (0,-6) {}; 
    \node[node label outer, anchor=north] at (f15.south) {$f_{15} = f_7^2$};
    
    \node[lattice node outer] (f16) at (6,-6) {}; 
    \node[node label outer, anchor=west] at (f16.east) {$f_{16} = f_8^2$};

    \draw[lattice line] (f4) -- (f5) -- (f6) -- (f7) -- (f8) -- (f1) -- (f2) -- (f3) -- cycle;
    \draw[lattice line] (f5) -- (f1);
    \draw[lattice line] (f3) -- (f7);
    \draw[lattice line] (f3) -- (f4);
    
    \draw[arrow line] (f2) -- (f6);
    \draw[arrow line] (f4) -- (f8);
    \draw[arrow line] (f1) -- (f5);
    \draw[arrow line] (f3) -- (f7);
    \draw[arrow line] (f0) -- (f1);
    \draw[arrow line] (f0) -- (f3);
    \draw[arrow line] (f0) -- (f2);
    \draw[arrow line] (f0) -- (f4);
    \draw[arrow line] (f0) -- (f6);
    \draw[arrow line] (f0) -- (f8);

    \draw[lattice line outer] (f10) -- (f11) -- (f12) -- (f13) -- (f14) -- (f15) -- (f16) -- (f9) -- cycle;
    \draw[lattice line outer] (f10) -- (f9);
    \draw[lattice line outer] (f12) -- (f14);
    \draw[lattice line outer] (f15) -- (f16);
    
    \draw[arrow line outer] (f2) -- (f10);
    \draw[arrow line outer] (f6) -- (f14);
    \draw[arrow line outer] (f4) -- (f12);
    \draw[arrow line outer] (f8) -- (f16);
    \draw[arrow line outer] (f1) -- (f9);
    \draw[arrow line outer] (f5) -- (f13);
    \draw[arrow line outer] (f3) -- (f11);
    \draw[arrow line outer] (f7) -- (f15);
    
\end{tikzpicture}
	\caption{A D2Q9 lattice (black lines) and a D2Q17 lattice (adding the red lines) configuration. For the OSSLBM model, the base configuration is D2Q9 with $f_{\alpha}^1$, and the additional $8$ velocities $f_{\alpha}^2$ are interpreted to carry data from the second time-level.}
	\label{fig:D2Q9 and D2Q17 lattice}
\end{figure}

The equilibrium functions $f_{\alpha}^{eq}(\bm{x},t)$ are expressed as explicit functions of the macroscopic variables:
\begin{equation}
    f_{\alpha}^{eq}(\bm{x},t) = {\cal{F}}_\alpha ( V_0, \bm{V_1})
    \label{eq:definition-distributions-from-macros}
\end{equation}
where $\cal{F}_\alpha$ contains all the information about the physics under consideration (hydrodynamics, acoustics, heat transfer etc.), leading to a fully general and versatile method.
A list of the different physics addressed in the present work with associated macroscopic quantities and equilibrium functions is displayed in Table~\ref{tab:physics-list}. Details are given in Appendix~\ref{sec:appendix-lbm}.
\begin{table}[!ht]
    \centering
    \caption{A list of possible physics models and related computational variables with associated equilibrium functions. The variables $h$, $\rho$ and  $\vu$ denote the water depth, the density and the velocity (depth-averaged velocity in the case of shallow water approximation), respectively, and $g$ is the gravitational acceleration. The subscript $0$ is related to base flow quantities which are inputs for the model. The weights $W_\alpha$ and the characteristic lattice speed $c_s$ are given in Table \ref{tab:lattice-features}.}
    \label{tab:physics-list}
    \begin{adjustbox}{angle=90}
    \begin{tabular}{c|c|c} 
       Case  & $(V_0, \bm{V}_1)$ & ${\cal{F}}_\alpha ( V_0, \bm{V_1})$ \\ \hline
        Low-Mach athermal flows & $(\rho, \rho \bm{u})$  & $ \displaystyle W_{\alpha}\left( V_0 + \frac{\bm{c}_{\alpha} \cdot \bm{V}_1}{c_s^2}+\frac{(\bm{c}_{\alpha} \cdot \bm{V}_1)^2}{2 V_0 c_s^4}-\frac{\bm{V}_1\cdot\bm{V}_1}{2V_0c_s^2}\right)$  \\ \hline
        Incompressible athermal flows & $(\rho', \rho_0 \bm{u})$ & $ \displaystyle W_{\alpha}\left( V_0 + \frac{\bm{c}_{\alpha} \cdot \bm{V}_1}{c_s^2}+\frac{(\bm{c}_{\alpha} \cdot \bm{V}_1)^2}{2 \rho_0 c_s^4}-\frac{\bm{V}_1\cdot\bm{V}_1}{2\rho_0c_s^2}\right)$ \\ \hline
        Linear acoustics &  $(\rho', \rho_0 \bm{u}')$ & $\displaystyle W_{\alpha}\left( V_0 C_0 + \frac{\bm{c}_{\alpha} \cdot \bm{V}_1}{c_s^2}+\frac{(\bm{c}_{\alpha} \cdot \bm{u}_0)(\bm{c}_{\alpha} \cdot \bm{V}_1)}{c_s^4}-\frac{\bm{u}_0\cdot\bm{V}_1}{\rho_0c_s^2}\right)$\\
        & & $\displaystyle C_0 = \left( 1 + \frac{\bm{c}_{\alpha} \cdot \bm{u}_0}{c_s^2}+\frac{(\bm{c}_{\alpha} \cdot \bm{u}_0)^2}{2 c_s^4}-\frac{\bm{u}_0\cdot\bm{u}_0}{c_s^2}\right)$ \\ \hline
        Shallow water & $(h, h\bm{u})$ & $\displaystyle  {\cal{F}}_0 = V_0 - \frac{5 g V_0^2}{6 c_s^2} - \frac{2 (\bm{V}_1 \cdot \bm{V}_1)}{3 V_0 c_s^2} $ \\
         & & $\displaystyle  {\cal{F}}_{\alpha=1,8} = D_\alpha \left(  \frac{g V_0^2}{6 c_s^2} + \frac{(\bm{c}_\alpha \cdot \bm{V}_1)}{3 c_s^2} + \frac{(\bm{c}_\alpha \cdot \bm{V}_1)^2}{2 V_0 c_s^4} - \frac{ (\bm{V}_1 \cdot \bm{V}_1)}{6 V_0 c_s^2}  \right) $ \\
         & & $D_{\alpha = 1,4} =1 , \, D_{\alpha = 4,8} = 1/4  $ \\ \hline
        Linearised shallow water & $(h', (h\bm{u})')$ & $ \displaystyle {\cal{F}}_0 = V_0 \left( 1 - \frac{5 g h_0}{c_s^2} + \frac{2 (\bm{u}_0 \cdot \bm{u}_0)}{c_s^2} \right) - \frac{4 (\bm{u}_0 \cdot \bm{V}_1)}{c_s^2}$ \\
          & &  $\displaystyle {\cal{F}}_{\alpha=1,8} = D_\alpha \left( V_0 C_0 +\frac{(\bm{c}_\alpha \cdot \bm{V}_1)}{3 c_s^2} + \frac{(\bm{c}_\alpha \cdot \bm{u}_0)(\bm{c}_\alpha \cdot \bm{V}_1)}{ c_s^4}  +\frac{(\bm{u}_0 \cdot \bm{V}_1)}{3 c_s^2} \right) $ \\
         & & $\displaystyle C_0 = \frac{g h_0}{3 c_s^2} - \frac{ (\bm{c}_\alpha \cdot \bm{u}_0)^2}{2 c_s^4} -  \frac{\bm{u}_0 \cdot \bm{u}_0}{6 c_s^2}$  \\
        & & $D_{\alpha = 1,4} =1 , \, D_{\alpha = 4,8} = 1/4  $ \\ 
    \end{tabular}
    \end{adjustbox}
\end{table}

\subsection{One-step simplified LBM}

Since the classical LBM relies on a collide-and-stream approach, it raises some encoding issues when dealing with quantum computing.
Therefore, an important step toward the derivation of a fully quantum LBM algorithm consists of reformulating the LBM in such a way that encoding will be made easier.
To this end, it is proposed here to address variants of LBM that do not rely on the collide-and-stream splitting, and more precisely to focus on single-step methods.

The final method used here, namely the one-step simplified LBM (OSSLBM) will be derived in two steps. First, a simplified LBM is obtained by considering evolution equations for macroscopic quantities, removing the equations for distribution functions (see Section~\ref{sec:SLBM}).
Second, the two steps are merged in a single one, yielding the final method (see Section~\ref{sec:qin}).

\subsubsection{Simplified LBM}
\label{sec:SLBM}

The first step toward designing a one-step simplified LBM is to switch to a simplified LBM. 
Here, the simplification means that the evolution equations for discrete distribution functions $f_{\alpha}$ are replaced by evolution equations for the macroscopic quantities $V_0$ and $\bm{V}_1$.
The key steps for deriving the simplified LBM are the following. Starting from the LBM-BGK model as in Eq.~\ref{eq:lbm-bgk}, one can recover the two following continuous evolution equations for macroscopic quantities:
\begin{equation}
    \frac{\partial V_0}{\partial t} + \nabla \cdot \left(\sum_{\alpha} \bm{c}_{\alpha}f_{\alpha}^{eq} \right) = 0,
    \label{eqn_8}
\end{equation}
\begin{equation}
    \frac{\partial  \bm{V}_1}{\partial t} + \nabla \cdot \Pi = 0, \quad \Pi_{ij} = \sum_{\alpha} \bm{c}_{\alpha,i}\bm{c}_{\alpha,j}\left( f_{\alpha}^{eq}+\left(1-\frac{1}{2\tau}\right)f_{\alpha}^{neq} \right).
    \label{eqn_9}
\end{equation}
The derivation entails a lengthy algebraic exercise in Taylor series expansion involving the zeroth- and first-order moments.

The non-equilibrium part is given by
\begin{equation}
    f^{neq}_{\alpha} = f_{\alpha} - f^{eq}_{\alpha} = -\tau\Delta t\left(\frac{\partial}{\partial t}+\bm{c}_{\alpha} \cdot \nabla\right)f_{\alpha}^{eq} = - \tau \Delta t  D_{\alpha} f_{\alpha}^{eq},
    \label{eqn_10}
\end{equation}
where the advection operator $D_\alpha$ is defined as
\begin{equation}
    D_{\alpha}=\left(\frac{\partial}{\partial t} + \bm{c}_{\alpha} \cdot \nabla\right).
    \label{eqn_11}
\end{equation}

Now, Eq.~\eqref{eqn_9} can be expanded at the leading order as
\begin{equation}
    D_{\alpha}f_{\alpha}^{eq}=-\left(1-\frac{1}{2\tau}\right)\nabla^2f_{\alpha}^{neq}-\frac{1}{\tau \Delta t}f_{\alpha}^{neq},
    \label{eqn_12}
\end{equation}
which can be recast as
\begin{equation}
 f_{\alpha}^{neq}= -\tau \Delta t \left(    D_{\alpha}f_{\alpha}^{eq}+\Delta t\left(\frac{1}{2}-\tau\right)D_{\alpha}^2f_{\alpha}^{eq} \right),
    \label{eqn_13}
\end{equation}
with a reminder that in the absence of external force one has
\begin{equation}
    \sum_{\alpha}f_{\alpha}^{neq} =  0, \quad \\ \sum_{\alpha}\bm{c}_{\alpha}f_{\alpha}^{neq} = 0.
    \label{eqn_14}
\end{equation}

This way, the right-hand side of Eq.~\eqref{eqn_11} can be expressed as a function of $f^{eq}_{\alpha}$ only, and there is no more need to solve evolution equations for the $f_{\alpha}$ since the equilibrium functions are functions of the sole macroscopic variables (see Table~\ref{tab:physics-list}).
The most popular version of the simplified LBM appears as a predictor-corrector scheme, see e.\,g. \cite{chen2017simplified,chen2017truly,chen2017simplifiedb,chen2018improvements,chen2023acoustic,chen2020simplified}.

\subsubsection{One-step SLBM -- combining the two steps}
\label{sec:qin}

As a further development on the simplified LBM, Qin et al. proposed to further simplify it by moving from a two-steps predictor-corrector to a single-step method, yielding a very efficient algorithm \cite{qin2022,qin2023a,qin2023novel,sikdar2023flexible}. 
This is achieved by computing the zeroth and first-order moments of Eq.~\eqref{eqn_13}, yielding:
\begin{equation}
    \sum_{\alpha} D_{\alpha}f_{\alpha}^{eq}+\Delta t\left(\frac{1}{2}-\tau\right)\sum_{\alpha} D_{\alpha}^2f_{\alpha}^{eq}=0,
    \label{eqn_15}
\end{equation}
and
\begin{equation}
    \sum_{\alpha} \bm{c}_{\alpha} D_{\alpha}f_{\alpha}^{eq}+\Delta t\left(\frac{1}{2}-\tau\right)\sum_{\alpha} \bm{c}_{\alpha} D_{\alpha}^2f_{\alpha}^{eq}=0.
    \label{eqn_16}
\end{equation}
Now using the following two second-order Taylor series expansions
\begin{equation}
    f_{\alpha}^{eq}(x\mp \bm{c}_{\alpha}\Delta t, t-\Delta t) = f_{\alpha}^{eq}(x,t) \mp \Delta t D_{\alpha} f_{\alpha}^{eq} + \frac{1}{2}\Delta t^2 D_{\alpha}^2 f_{\alpha}^{eq} + O(\Delta^3),
    \label{eqn_17}
\end{equation}
and
\begin{equation}
    f_{\alpha}^{eq}(\bm{x}\mp 2\bm{c}_{\alpha}\Delta t, t-2\Delta t) = f_{\alpha}^{eq}(\bm{x},t) \mp 2\Delta t D_{\alpha} f_{\alpha}^{eq} + 2\Delta t^2 D_{\alpha}^2 f_{\alpha}^{eq} + O(\Delta^3),
    \label{eqn_18}
\end{equation}
the authors derived some linear combinations that approximate Eqs.~\eqref{eqn_17} and \eqref{eqn_18}, which can be recast as
\begin{equation}
    V_0 (\bm{x},t) = \frac{\tau-1}{2-\tau}\sum_{\alpha}f_{\alpha}^{eq}(\bm{x}-2\bm{c}_{\alpha}\Delta t, t - 2\Delta t) + \frac{3-2\tau}{2-\tau} \sum_{\alpha}f_{\alpha}^{eq}(\bm{x}-\bm{c}_{\alpha}\Delta t, t-\Delta t),
    \label{eq:macros-oss-density}
\end{equation}
\begin{equation}
    \bm{V}_1 (\bm{x},t) = \frac{\tau-1}{2-\tau}\sum_{\alpha}\bm{c}_{\alpha}f_{\alpha}^{eq}(\bm{x}-2\bm{c}_{\alpha}\Delta t, t - 2\Delta t) + \frac{3-2\tau}{2-\tau} \sum_{\alpha}\bm{c}_{\alpha} f_{\alpha}^{eq}(\bm{x}-\bm{c}_{\alpha}\Delta t, t-\Delta t).
    \label{eq:macros-oss-momentum}
\end{equation}

This scheme relies on a D2Q17-like lattice (Figure~\ref{fig:D2Q9 and D2Q17 lattice}), with the interpretation that nodes are interpreted on two different time-levels: $(t-\Delta t)$ matching with $f_{\alpha}^1$ and $(t-2\Delta t)$ with $f_{\alpha}^2$.  We will consider this as a basis for the development of the quantum algorithm for LBM, with the so-called Tau1 method as a special cases.

The main benefit of the OSSLBM model is that it allows parametrization beyond the simplification Tau1. The addition of a second time-level brings relatively little computational overhead on a quantum system, as the operators can be built to act on both time-levels in superposition. Additionally, since the main parts of the algorithm work directly with the macroscopic field variables instead of distribution functions, handling for example boundary conditions is easier and more direct.

It was shown in \cite{qin2023novel} that this method achieves the second order of accuracy, which makes it more accurate than a previous scheme with a single time step storage requirement (see \cite{delgado2021single,delgado2022efficient}) and simpler than more sophisticated schemes like \cite{liu2024macroscopic}.

\subsubsection{The Tau1 method}
\label{sec:tau1}

The Tau1 method \cite{bacza2023latticeBoltzmannmethodfluid,matyka2021,zhou2021} is a long-known method referred to as \quotes{collisionless LBM}, \quotes{macroscopic LBM}, or \quotes{one-step SLBM}. Setting $\tau = 1$ in Eq.~\eqref{eq:discrete-boltzmann} becomes
\begin{equation}
    \frac{\partial f_{\alpha}}{\partial t} + \bm{c}_{\alpha} \cdot \nabla f_{\alpha} = - \left(f_{\alpha}(\bm{x},t) - f^{eq}_{\alpha}(\bm{x},t)\right)
    \label{eqn_21}
\end{equation}
while the LBM in Eq.~\eqref{eq:lbm-bgk} degenerates as follows
\begin{equation}
    f_{\alpha}(\bm{x}+\bm{c}_{\alpha}\Delta t, t+\Delta t) = f^{eq}_{\alpha}(\bm{x},t),
    \label{eqn_22}
\end{equation}
showing that the classical LBM simplifies as a pure streaming step, which corresponds to first
order accuracy in both space and time explicit Euler or 1st order upwind scheme discretisation for the advection operator $D_i$. 

This discretization implicitly introduces a dissipation that is a function of $\Delta t$, $\nu=c_s^2\Delta t/2$, explaining why the effective viscosity can be tuned by selecting $\Delta x$ (since $\Delta t$ is enslaved to $\Delta x$ in LBM). An explicit control of the viscosity is proposed in \cite{zhou2021} via manipulation of the magnitudes of the discrete velocities. Instead of considering discrete velocities with components equal to 0 or $\pm 1$, the author considers $\boldsymbol{c}_i$ with components equal to $\{0,\pm e\}$, where
\begin{equation*}
    e=6\nu/\Delta x, c_s=e \Leftrightarrow \nu = e\Delta x/6=e^2\Delta t/6,
\end{equation*}
where $\nu$ is the fluid viscosity to be enforced. The time step is then tuned so that $\Delta t=\Delta x/e=\Delta x^2/6\nu$ so that there is no need to interpolate data.

Another way to recover explicit control over viscosity would be to use a lattice kinetic scheme and its sequels~\cite{inamuro2002lattice,suzuki2014improved,suzuki2021simple}, which can be seen as a linear modification of the Tau1 method that introduces a parameter governing the viscosity explicitly. This family of schemes was not retained to ease the development of the quantum algorithm. The reason is that merging it with the LBM requires several additional steps, since it also modifies the value of $c_s$ (and then the Mach number), leading to inconsistencies with LBM.

Taking the zeroth- and first-order moments of Eq.~\eqref{eqn_22}, one obtains
\begin{equation}
    V_0 (\bm{x},t) = \sum_{\alpha}f_{\alpha}^{eq}(\bm{x}-\bm{c}_{\alpha}\Delta t, t-\Delta t),
    \label{eqn_23}
\end{equation}
\begin{equation}
     \bm{V}_1 (\bm{x},t) = \sum_{\alpha}\bm{c}_{\alpha} f_{\alpha}^{eq}(\bm{x}-\bm{c}_{\alpha}\Delta t, t-\Delta t).
    \label{eqn_24}
\end{equation}
One can also check that these expression are recovered by taking $\tau = 1$ in the OSSLBM, ensuring the global consistency of the three methods.

The Tau1 method is also straightforwardly recovered from the second order of the OSSLBM by considering the following second-order accurate space-time extrapolation along characteristic lines related to discrete velocities $\bm{c}_\alpha$:
\begin{equation}
f_\alpha^{eq} (\bm{x} - 2 \bm{c}_\alpha \Delta t, t - 2 \Delta t) = 2  f_\alpha^{eq} (\bm{x} -  \bm{c}_\alpha \Delta t, t -  \Delta t) - f_\alpha^{eq} (\bm{x} , t)
\label{eqn_25}
\end{equation}
It is worth noting that the Tau1 method is the one considered by many researchers when dealing with the derivation of a quantum algorithm for the LBM, since it allows for a drastic simplification of the collision term in Eq.~\eqref{eq:collision}~\cite{budinski2021,budinski2022,kocherla2024two,wawrzyniak2025dynamic,tiwari2025algorithmic}.

\subsection{Boundary conditions} 
\label{sec:intro-boundary}

A major difference between the standard LBM and OSSLBM is that the latter directly works on the macroscopic variables.
Consequently, boundary conditions are defined on the macroscopic density and momentum terms as in the usual methods based on the Navier-Stokes equations, without the need to consider equilibrium functions.
Therefore, standard combinations of Dirichlet and Neumann conditions for the density and the momentum can be used to model inlet, outlet, no-slip and slip boundary conditions.
In order to facilitate the transfer to a quantum setting, Neumann conditions are implemented using first-order accurate schemes.

Due to the use of a D2Q17-like stencil, the OSSLBM cannot be used in the first layer of cells near domain boundaries.
Therefore, it should be replaced by another method with a narrower stencil. In the present implementation, it is replaced by the Tau1 method in cells in which the OSSLBM can not be used -- this means that the boundary conditions are primarily applied on the first time-level only, with additional rebalancing to recover the correct scaling for Eqs.~(\ref{eq:macros-oss-density}--\ref{eq:macros-oss-momentum}).

It is worth noting that, considering Eq.~\eqref{eqn_25}, using the Tau1 method in the first layer of cells can be interpreted as using the OSSLBM in combination with an extrapolation technique to reconstruct missing data at virtual nodes located outside the computational domain. 

\subsection{Masking method for immersed solid bodies} 
\label{sec:intro-maskingmethod}

Since the implementation of complex geometries is made much more difficult by quantum implementation, it is proposed to use a very simple method to account for arbitrary immersed solid bodies.

In the present implementation, solid bodies are modeled by a masking method that sets to zero all fluctuations in cells located inside the bodies.
This approach belongs to the class of volume penalization methods~\cite{angot1999penalization,llorente2024theoretical,zhdanova2022generalized} and fictitious domain approaches~\cite{glowinski2001fictitious}.
More precisely, it mimics the macroscopic Navier-Stokes-Brinkmann equations, with  infinite relaxation parameters toward a medium a rest inside solid bodies.
This approach has been assessed for both steady and moving obstacles for several numerical methods applied to Navier-Stokes equations, ranging from finite differences to spectral methods, see  e.\,g. \cite{khadra2000fictitious,kolomenskiy2009Fourier,kadoch2012volume}.

\section{Quantum algorithm for LBM}%
\label{sec:quantumlbm}

In order to formulate the OSSLBM as a quantum algorithm, one would hope to transform the algebraic expressions into quantum circuits and gates. However, this process is not quite that straightforward due to the nature of quantum computing in general and qubit circuit systems in particular. For this reason, we do not aim to define a true one-to-one correspondence between a classical and a quantum algorithm, but rather present a more abstract way of designing quantum routines in such a way that the overall structure pairs with the OSSLBM. Most of the subroutines here are generic in the sense that they can be used beyond the OSSLBM and even the LBM in general. Therefore, there will be explicit differences in how we name for example variables and operators.

In this section, we analyse separately the quantum subroutines representing parts of the LBM algorithm. We focus on the circuit blocks for one time-step, and describe the full simulation process later in Section~\ref{sec:analysis}. In search for greater efficiency in each section of the algorithm, we study different techniques for generating the corresponding subcircuits instead of attempting to generate a single unitary operator encompassing every segment. The examples we give for the subcircuits are written in such a way that they can be directly linked together to form the total circuit representing one time-step of a simulation.

\subsubsection*{Lattice configurations}

In the following description, we study two primary LBM models: the full OSSLBM as in Section~\ref{sec:qin}, and the Tau1 simplification (Section~\ref{sec:tau1}). Our example lattice configurations are 2D and 3D models with prescribed discrete velocities, where we use the definitions described in Table~\ref{tab:lattice-examples}.
\begin{table}[!htb]
    \caption{Example lattice configurations used for the quantum algorithm.}
    \label{tab:lattice-examples}
    \centering
    \begin{tabular}{p{.25\textwidth}cc}
        \toprule
                                    & 2D        & 3D \\
        \cmidrule{2-3}
        {OSSLBM \newline
        two time-levels}
                                    &  D2Q17    & D3Q53 \\
        \midrule
        {Tau1  \newline              
        one time-level}             
                                    &  D2Q9     & D3Q27 \\
        \bottomrule
    \end{tabular} 
\end{table}

Note especially that we will always assume that the one time-level model follows the Tau1 construction. In literature one can find more general models attached to the D2Q9 and D3Q27 lattice configurations, but we will make the restriction to Tau1. 
The second time-level in the full model effectively doubles the number of the discrete non-zero velocities, although the evolutions of these two sets are independent of each other from the perspective of the collision and the streaming operators.

The models may combine both the Tau1 and the full scheme. In particular, it may be helpful to use the Tau1 scheme for some initial time-steps to stabilize the model, and then switch to the full set to gain better accuracy and flexibility in the parametrization.

\subsection{Overview of the structure}

\begin{figure}[!htb]
	\centering
	\begin{tikzpicture}[
    block/.style={rectangle, draw, text width=2.5cm, text centered, rounded corners, minimum height=1cm},
    arrow/.style={-Latex, thick},
    label/.style={text width=1.5cm, text centered, font=\small}
]

\node (initial) at (0,2) {Initialization};
\node[block] (collision) at (0,0) {Collision};
\node[block] (propagation) at (4,0) {Propagation};
\node[block] (integration) at (8,0) {Integration};
\node[block] (boundary) at (12,0) {Boundary};
\node (readout) at (12,-2) {Readout/iteration};

\draw[arrow] (initial) -- (collision);
\draw[arrow] (collision) -- (propagation);
\draw[arrow] (propagation) -- (integration);
\draw[arrow] (integration) -- (boundary);
\draw[arrow] (boundary) -- (readout);

\end{tikzpicture}
	\caption{General representation of the quantum algorithm for the OSSLBM.}
	\label{fig:overview-algorithm}
\end{figure}
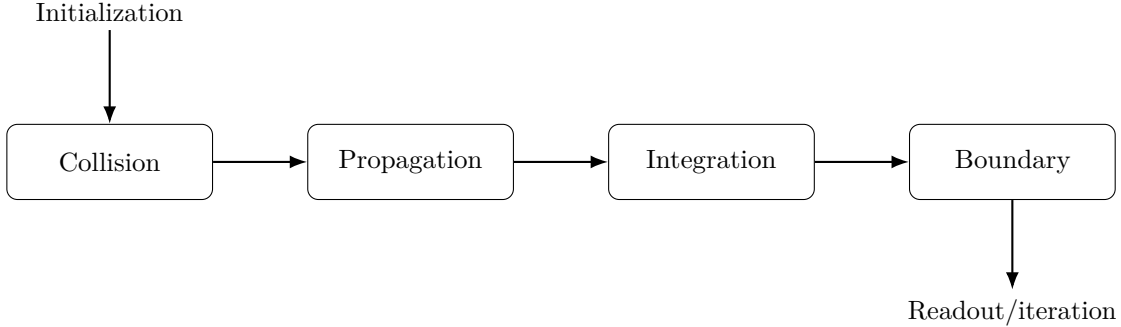

The quantum algorithm is directly determined by the structure of the OSSLBM, and can be defined through four major steps (see Figure~\ref{fig:overview-algorithm}): collision, propagation, integration, and enforcement of boundary conditions.  Note that these steps may not exactly map to what is usually considered as parts of a LBM model, due to the different structure needed to create the quantum-computational routines.

An initial quantum state is generated from the macroscopic lattice fields $\rho$ and $\vu$ at a given $t_0$. We use generic terms \emph{density} and \emph{momentum} fields for $\rho$ and $\vu$: these are to be understood as within-the-model placeholders, and interpretations to concrete physics are left up to the case (in general, these variables should match with $V_0$ and $\mathbf{V}_1$ as used in Section~\ref{sec:lbm}). The assumption is that we can build an LBM model with a pair of fields consisting of a (pseudo)scalar and a vector field -- however, the basis could be more complex, such as having also nonlinear terms arising from these fields as in Section~\ref{sec:collision-nonlinear}.

The computational basis states over the lattice register are used to encode the lattice position, and the field values are mapped into the amplitudes. For an efficient encoding method, we can employ for example \cite{kerppo2025} and similar techniques; in general, encoding is a problem in its own, and we will not discuss it beyond a few examples here. Unlike in some earlier quantum LBM setups~\cite{budinski2021, budinski2022}, the amplitudes are not required to be positive, and it is conceivable that they could also be complex in some physics models.

A collision-like operator is introduced to calculate the equilibrium distribution functions $f_{\alpha}^{eq}$. In general, this is based on Eq.~\eqref{eq:definition-distributions-from-macros}, with details depending on the exact physics. In what follows, we call these simply \emph{distribution functions} and denote them by $f_{\alpha}^k$, where $k\in\{1,2\}$ refers the time-level specific subsets used in the full OSSLBM model.

A propagation operator shifts (spatially and temporally) the distribution data along local lattice velocities as in Fig.~\ref{fig:D2Q9 and D2Q17 lattice}. The propagation operator connects the local collision dynamics to the evolution of the global lattice state by linking neighbouring lattice sites to each other.

From the propagated distributions the macroscopic fields are then reconstructed by calculating the zeroth and first statistical moments, following Eqs.~(\ref{eq:macros-oss-density}--\ref{eq:macros-oss-momentum}). This is derived from the local integration of the distributions, and boils down to simple sums due to the lattice discretization.

A boundary operator enforces external boundary conditions along the solid domain boundaries. One can then further simulate an immersion of an object in the domain using the masking method (as in Section~\ref{sec:intro-maskingmethod}): we call this the object operator. In Fig.~\ref{fig:overview-algorithm}, the \emph{Boundary} operator includes also the possible object operator as an enforcement of internal boundary conditions.

The final state then carries information on the new macroscopic fields, which can used to launch the next time step, or prepared for the final measurements. More details on the time-stepping can be found in Section~\ref{sec:analysis}.

To aid in exposition on a more concrete level, we introduce three quantum registers: the lattice register $q$, the superposition register $s$, and the computational ancilla register $a$. %
The size of the lattice register $q$ is determined by the number $N$ of lattice sites: the number of qubits being $n(q) = \log_2 N$. %
The superposition register $s$ is used for storing the field values over the lattice, where the number of qubits $n(s)$ depends primarily on the number of spatial dimensions of the problem being simulated: for example, the D2Q9 lattice requires at least four qubits to store the nine distributions in superposition over the lattice. %
The size of the ancilla register $a$ depends on the level of circuit optimization and the dimensions of the problem; see details in Section~\ref{sec:analysis}. In general, the number of qubits in the register $a$ can be fixed to a relatively low number independent of the size of the lattice.

We propose several new techniques for finding efficient sequences of quantum gates that implement the desired operators. The exact methods depend on the structure of the given operator, the dimensions of the problem as well on the type of the OSSLBM being used (full model or Tau1).
In terms of efficiency, we have two main goals: the computational complexity measured in qubit and gate resources needed, and the minimization of the residual states stemming from the non-unitary operations inherent in the LBM models. Reducing the amount and the amplitude of the residual substates will significantly impact the desired probabilities during the final measurement. More on this in Section~\ref{sec:analysis}.

\subsection{Managing the two time-levels}

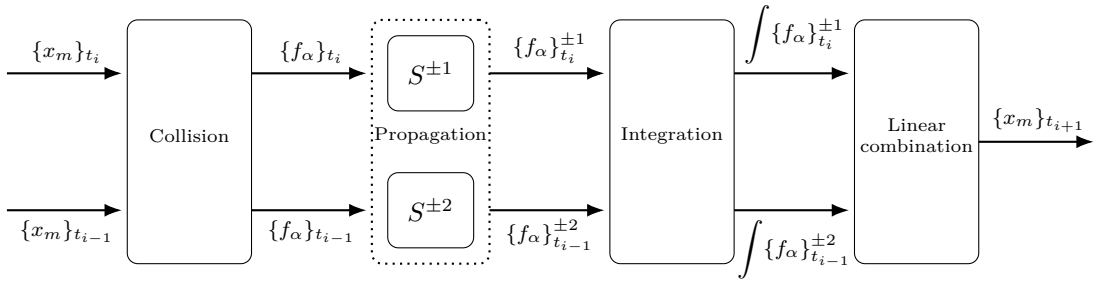
\begin{figure}[!htb]
	\centering
\begin{tikzpicture}[
    node distance = 4mm and 16mm, 
    block/.style={rectangle, draw, text centered, rounded corners, minimum height=1cm, minimum width=3.25em, inner sep=4pt, outer sep=6pt},
    block_invisible/.style={block, opacity=0, minimum width=3.5em},
    block_endpoint/.style={block_invisible, minimum width=0em},
    block_surround/.style={block, inner sep=0pt, outer sep=0pt, font={\scriptsize}},
    arrow/.style={-{Latex[sep=2pt]}, thick, font={\footnotesize}}
]

\node[block_endpoint] (init_one) at (0,0) {};
\node[block_endpoint, below = of init_one] (init_two) {};

\node[block_invisible, right = of init_one] (collision_one) {};
\node[block_invisible, below = of collision_one] (collision_two) {};
\node[block_surround, fit=(collision_one) (collision_two)] (colblock) {Collision};

\node[block, right = of collision_one] (shift_one) {$S^{\pm1}$};
\node[block, right = of collision_two] (shift_two) {$S^{\pm2}$};
\node[block_surround, dotted, thick, fit=(shift_one) (shift_two)] (propblock) {Propagation};

\node[block_invisible, right = of shift_one] (integration_one) {};
\node[block_invisible, right = of shift_two] (integration_two) {};
\node[block_surround, fit=(integration_one) (integration_two)] (integblock) {Integration};

\node[block_invisible, right = of integration_one] (combine_one) {};
\node[block_invisible, right = of integration_two] (combine_two) {};
\node (combine) [block_surround, fit=(combine_one) (combine_two)] {Linear\\combination};

\node[block_endpoint, right = of combine] (final) {};

\draw[arrow] (init_one) -- node[above] {$\{x_m\}_{t_i}$} (collision_one);
\draw[arrow] (init_two) -- node[below] {$\{x_m\}_{t_{i-1}}$} (collision_two);

\draw[arrow] (collision_one) -- node[above] {$\{f_{\alpha}\}_{t_i}$} (shift_one);
\draw[arrow] (collision_two) --  node[below] {$\{f_{\alpha}\}_{t_{i-1}}$} (shift_two);

\draw[arrow] (shift_one) -- node[above] {$\{f_{\alpha}\}_{t_i}^{\pm1}$} (integration_one);
\draw[arrow] (shift_two) --  node[below] {$\{f_{\alpha}\}_{t_{i-1}}^{\pm2}$} (integration_two);

\draw[arrow] (integration_one) -- node[above] {$\displaystyle\int\{f_{\alpha}\}_{t_i}^{\pm1}$} (combine_one);
\draw[arrow] (integration_two) -- node[below] {$\displaystyle\int\{f_{\alpha}\}_{t_{i-1}}^{\pm2}$} (combine_two);

\draw[arrow] (combine) -- node[above] {$\{x_m\}_{t_{i+1}}$} (final);

\end{tikzpicture}
	\caption{The evolution of the two time-levels in the algorithm. The collision and integration operators act on the two levels in parallel (full superposition), while the propagation operator assigns them different shift steps (partial superposition). Here $\{x_m\}_{t_i}$ refers to the generic set of field variables at a given time $t_i$; these could be for example the pair $(\rho, \vu )$.}
	\label{fig:overview-timelevels}
\end{figure}

The OSS model for the LBM requires managing two sets of variables corresponding to two discrete time-levels. To update the field variables from the time $t_i$ to the time $t_{i+1}$, one needs the field data also from the time $t_{i-1}$. This two-level evolution is illustrated in Fig.~\ref{fig:overview-timelevels}. The main difference is in how the distribution functions are propagated: at the time-level $t_i$, the propagation step is $\pm1$ on the lattice, while at the time-level $t_{i-1}$ the propagation step is $\pm2$. The propagated distributions are then locally integrated (separately at each level) before combining the levels according to the Eqs.~(\ref{eq:macros-oss-density}--\ref{eq:macros-oss-momentum}).

Keeping track of the variables from different time-levels requires to carry a \quotes{copy} of the previous computation results in some substate, as there is no quantum RAM for storage.%
~\footnote{By copy, we do not mean copying a quantum state, but rather that a second set of the initial amplitudes is retained in some substate (subject to global normalization). For example, an array of amplitudes over qubits $(q_0, \dots, q_n)$ could be \quotes{copied} in this sense with another qubit $q_{n+1}$ by placing an $\hgate$-gate on it. All the amplitudes are changed by this, but can be regained through renormalization.} 
This is reflected in the collision operator (see the matrix in Eq.~\eqref{eq:c-matrix}), as well as in the manipulation of the $s$-register substates so that the copies can be kept intact during the computation.

\subsection{Collision and local equilibrium distributions}\label{sec:collision}

The collision operator maps the macroscopic fields over the lattice to (local) velocity-related distributions. This mimics the collision in cellular automata schemes such as the lattice gas automata, where particles are assigned explicit collision rules at each lattice site. In the standard LBM scheme, this collision rule is played on the distribution functions instead and formally defined as a collision integral that models the interaction between distributions of different velocities. Here, we retain the term \emph{collision} for the operator that generates the propagating distribution functions, even though in our case the map is not between distributions but from lattice fields to distributions.

This operator is used to calculate the distribution functions $f_{\alpha}^1$ and $f_{\alpha}^2$ from the macroscopic fields 
$\rho$ and $\vu$, where the field data includes both time levels.
The equations for the distribution functions $f^k_{\alpha}$ depend on the type of the physical process that we are simulating. For example, if we are solving the Navier-Stokes equations, the collision operator is built from the Eq.~\eqref{eq:navier-stokes}, while in the case of the acoustic wave propagation the equation would be Eq.~\eqref{eq:acoustics}. We make a distinction between the \emph{linear} and the \emph{nonlinear} models, focusing first on the linear collision model.

Linear collision operators can be described as a matrix acting on a vector encoding the macroscopic field data on a given basis.
The locality of the collision step implies that the operator has dimensions fixed by the lattice configuration, and with the entries corresponding to the particular physical process (distribution functions). If the vector containing the amplitude-encoded field data at one lattice site $x$ is denoted as $\mathbf{m}$ and the equilibrium matrix as $C$, the distribution functions at one lattice site are given by 
$C : \mathbf{m}(x,t) \mapsto \mathbf{f}$,
where the matrix $C$ has the block form 
\begin{equation}
    C = 
    \begin{bmatrix}
        W & 0 \\
        I & 0 \\
        0 & 0
    \end{bmatrix} .
    \label{eq:c-matrix}
\end{equation}
Here, the block $W$ contains real weight coefficients dictated by model physics and the lattice configuration (related to Table~\ref{tab:lattice-features}). Crucially, these weights are \emph{independent of the chosen lattice site}, and we can apply the operator over the superposition of states encoding the lattice. Typically, the weights are also considered to be independent of the time $t$.

The sizes of the blocks $W$ and $I$ depend on the lattice model and the type of the problem, the number of rows in $W$ matching with the number of discrete lattice velocities. For example, the number of rows in $W$ is $9$ and $27$ for D2Q9 and D3Q27 models, respectively. For linear problems we typically use the input vector to encode variables $\rho$ and $\vu$, where the number of components of the momentum field $\vu$ depends on the space dimension -- consequently, both $W$ and $I$ would have rank three for D2Q9 and rank four for D3Q27 models.
The identity block matches the dimension of the unpadded input vector in order to carry through a copy of the initial state for the next time step in the models with two time-levels.

A quantum circuit embedding of the matrix requires that its total dimension matches at least the minimum power of $2$ needed to include all these coefficients. For example, $\dim(C) = 16$ for the D2Q9 model, and $\dim(C)=32$ for the D3Q27 model, and the remaining blocks are padded with zero. If necessary, the rows of the matrix can be freely reordered. This can be helpful in simplifying the circuit for the following propagation step.

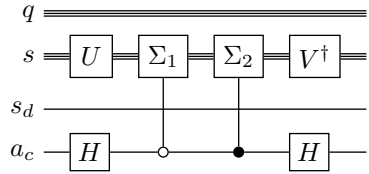
\begin{figure}[htb]
	\centering
\begin{tikzpicture}
    \begin{yquant}

    qubits {$q$} q;
    qubits {$s$} s;
    qubit {$s_d$} sd;
    qubit {$a_c$} a;

    box {$U$} s;

    h a;
    box {$\Sigma_1$} s | ~a;
    box {$\Sigma_2$} s | a;
    h a;
    
    box {$\dagg{V}$} s;

    \end{yquant}
\end{tikzpicture}

	\caption{An example circuit implementation for the collision step using the SVD to decompose the collision matrix. The size of the register $s$ depends on the model used. In any case, the last qubit of the register here marked with $s_d$ will be used to stored the variables relating to the second time-level. The collision operator then acts on both time-levels in superposition.}
	\label{fig:Quantum circuit for the equilibrium step}
\end{figure}

Given that $C$ is in general non-unitary, it requires a unitary embedding for the quantum circuit synthesis. We can use a factorization of the non-unitary operator into a linear combination of unitaries. 
As a concrete factorization example here, we use singular value decomposition (SVD). This allows us to decompose the matrix into a product of scaling and rotational matrices in the form $C = U\Sigma\dagg{V}$, where $U$ and $V$ are the unitary rotation matrices and the scaling matrix $\Sigma$ is diagonal. The burden of embedding is now shifted on the the diagonal matrix $\Sigma$, which in general is not unitary. For example, we can apply an additional decomposition of the matrix $\Sigma$ into a linear combination of two unitary operators $\Sigma_1$ and $\Sigma_2$ following \cite{xin2020}, where each unitary has a weight factor of $0.5$. This allows a circuit implementation of the embedding very efficiently just by using two Hadamard gates on the control qubit. Figure~\ref{fig:Quantum circuit for the equilibrium step} depicts such a quantum circuit implementation for the total embedding of the matrix.

Depending on the rank of the matrix $C$, it may be possible to use reduced SVD matrices that act only on a part of the register $s$. For example, given an input vector with just the density $\rho$ and the components of the momentum $\vu$, the rank is at most four. The matrices $U$ and $\Sigma$ can be then reduced to a size $4\times4$. However, the operator $\dagg{V}$ needs to always act on the full register $s$.

Other possible choices for the decomposition of a non-unitary matrix include Pauli decomposition procedure~\cite{hantzko2024} as well as unitary decomposition technique used in \cite{xin2020,budinski2021}. 
However, the SVD decomposition is a particularly good option due to the simplicity of the circuits. It is important here to note that while different matrices lead to slightly different circuits at the gate level, the described process for the collision step is fundamentally the same regardless of the type of the model -- namely, whether one is implementing the full or the Tau1 model. The collision operator acts only on the superposition register, so that the calculation of the distribution functions across the lattice and for both time-levels follows from the superposition.

In Tau1 models we are only interested in the time level $t-\Delta t$ (see Eqs.xxxxx). In that case the calculation of the distributions functions $f_{\alpha}$ requires only the $t-\Delta t$ time level, and we do not need to carry a copy of the state throughout. This means that additional reduction of the matrix may be possible: for example, the size of the operators for the 2D models could be reduced, and consequently the embedding itself would not need to act on the full $s$ register.%
~\footnote{Some additional entangling of the total $s$ register is still required to create the full distribution set; the hope is that this entangling is less costly than the full embedding would be.} %
This requires that one of the  rows in the block $W$ is linearly dependent on the others, so that such a reduction can be made by tweaking the encoding and the matrix entries. 

\subsubsection{Amplitude dissipation and unitary collision operators}

The generic collision matrix from Eq.~\eqref{eq:c-matrix} is non-unitary. Since this requires unitary embedding for the diagonal $\Sigma$, some dissipation of the amplitudes is expected. In the embedding, we have the following recipe for the operator $\Sigma_1$:
\begin{equation}
    \Sigma_1(k,k) = \sigma_k + j \sqrt{1 - \sigma_k^2} ,
\end{equation}
where $\sigma_k$ are the singular values of the matrix $C$ indexed in descending order. This, combined with $\Sigma_2 = \dagg{\Sigma}_1$, gives us some sense of the dissipation. Strong deviation of the singular values from the maximum value $\sigma_1$ indicates higher dissipation, as well as sensitivity of the amount of dissipation with respect to the input state. Similarly, zero singular values will contribute to the dissipation, which can be a problem when the rank of matrix $C$ is low.

Since the embedding error is coherent, some amplitude amplification scheme could be used to reduce the dissipation. We will not analyze this further here, but leave it as an implementation detail for future study.

It may be also possible to form a unitary matrix that fulfills the required equations. This hinges on the fact that most of the entries in the amplitude vector would be zero for a given model, and that the weights dictated by the lattice model are suitable. In particular, this requires symmetry in the choice of lattice velocities, so that the full matrix can be orthogonalized. Following Table~\ref{tab:lattice-features}, the 1D and 2D lattice configurations are projections of the 3D models~\cite[Ch.~3.4.7]{kruger2016lattice}, which implies possibilities for dimensional decomposition.
For example, the nonreduced linear collision matrix in the D2Q9 model is of size $16\times16$, and it acts on a vector containing only three potentially non-zero scalars $\rho(x)$, $u_1(x)$ and $u_2(x)$ at each site $x$, the remaining of the entries in the vector being zero. This would allow us to fill the zero entries in the matrix in such a way that the operator can become unitary without changing the outcome on the states of this particular form. The obvious benefit in doing this is the straightforward ancilla-free circuit embedding with no dissipation of the amplitudes.

\subsubsection{Nonlinear collision models}%
\label{sec:collision-nonlinear}

For nonlinear collision models, we consider here a hybrid computation loop, where the nonlinear terms are built during classical pre-processing. One can then construct a similar matrix as with the linear case, this time acting on a statevector that encodes both the linear and the nonlinear terms. For example, the D2Q9 Navier-Stokes model requires an input vector $\mathbf{m}$ of at least dimension six (in terms of fields included), the terms being the following:
\begin{equation}
    \{x_m\} = \{ \rho, u_1, u_2, u_1^2, u_2^2, u_1u_2 \} .
    \label{eq:nonlinear-terms}
\end{equation}
From this data one can construct a matrix operator that fulfills the quadratic expansion given by Eq.~\eqref{eq:navier-stokes}.

The quest for a purely quantum implementation of the nonlinear terms is an actively studied research topic. Interesting candidates to mention involve either Carleman linearization~\cite{bastidazamora2026} or machine learning techniques~\cite{corbetta2023LBcollision,lacatus2025surrogatequantumcircuitdesign,Itani2025QMLLBM,rbf2-p8tf}. We will not discuss these approaches in more detail in this study; the example results for nonlinear models in Section~\ref{sec:example_airfoil} use the hybrid approach outlined above.

\subsection{Propagation of distribution functions}

Following the cellular automata logic underlying the LBM, the distribution functions gleaned from the collision are propagated (or streamed) along the prescribed local lattice velocities. The propagation spreads the local dynamics across lattice sites, thus enabling the evolution of the global state. As an example, the local velocity configuration for D2Q9/Q17 model is illustrated in Fig.~\ref{fig:D2Q9 and D2Q17 lattice}, and this holds equally for all the sites.

The quantum routine for propagation  is essentially a quantum walk, where the superposition of the directions is determined by the collection of distributions functions. The lattice configuration dictates how many links there are in the lattice: below we will outline the D2Q9 configuration as an example. Additionally, there are several different ways to construct the propagation sequence. We present here a variation where each lattice dimension is propagated separately. Although this may not always be the most efficient method, it provides a good example due to its structural simplicity. Since in this step we are not interested in generating the walk distribution itself, we will call these routines \emph{shifts} instead of walks. See \cite{shift-preprint} for more details on how shift patterns can be constructed over a regular lattice. Note that the shift as prescribed here is \emph{periodic} by construction. This can be used as a built-in boundary condition if periodicity is desired.

For an $n$-dimensional orthogonal lattice, we have essentially two different types of propagation directions. The simple ones relate to a single dimension,  and are therefore orthogonal and separable. We assume that such directions are always present. Additionally, we may have directions that combine dimensions, or apply a \emph{diagonal} propagation on the lattice. For example, the D2Q9 lattice admits four distinct orthogonal directions and four diagonal directions, aside the last \emph{stationary} (zero local velocity) direction. Whereas the orthogonal directions can be separated and the associated distributions propagated via single shift routine, the distributions associated with the diagonals need to be acted upon by a combined shift of all the relevant dimensions.

The general scheme is the following. First, we group together the distribution function substates associated to a given dimension, and act upon those substates with the correct shift routine. We then regroup the distribution function substates according to another dimension, and act upon the regrouped substates with the shift routine pertaining to that dimension. This repeats as many times necessary to fulfill the lattice configuration. Additional regroupings may be done before and after to link the propagation subcircuit to the preceding collision, and to the integration step after.

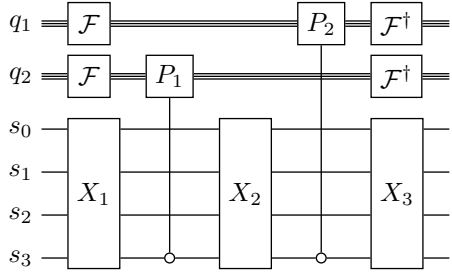
\begin{figure}[!htb]
    \centering
\begin{tikzpicture}
    \begin{yquant}
        qubits {$q_1$} q1;
        qubits {$q_2$} q2;
        qubit s[4];

        box {$X_1$} (s);

        box {$\mathcal{F}$} (q1);
        box {$\mathcal{F}$} (q2);

        box {$P_1$} (q2) | ~s[3];

        box {$X_2$} (s);

        box {$P_2$} (q1) | ~s[3];

        box {$\dagg{\mathcal{F}}$} (q1);
        box {$\dagg{\mathcal{F}}$} (q2);

        box {$X_3$} (s);
    \end{yquant}
\end{tikzpicture}
    \caption{An example implementation for the D2Q9 propagation, where the lattice register is split into $q_1$ (horizontal dimension) and $q_2$ (vertical dimension). The blocks with $\mathcal{F}$ denote the quantum Fourier transform. We use the flag qubit state $s_3 = 0$ to control the two shifts, and reordering operators $X_k$ to shuffle the distribution date accordingly. One can extend this scheme to the D2Q17 model by simply adding extra phase gate blocks ($P_n$) controlled by an additional qubit $s_4$, which is used to store the data $f_{\alpha}^2$ for the second time level. Note that the first reordering operator $X_1$ can be at least partially embedded as a row-ordering of the equilibrium matrix $C$ in the preceeding collision step. }
    \label{fig:circuit-propagation-d2q9}
\end{figure}

For the D2Q9 lattice example, we need two shifts: the vertical and the horizontal. The set of distribution functions associated with the vertical directions according to the Fig.~\ref{fig:D2Q9 and D2Q17 lattice} is $\{f_2, f_3, f_4, f_6, f_7, f_8\}$, where $f_3$ and $f_7$ are purely vertical. Similarly, the set of distribution functions $\{f_1, f_2, f_4, f_5, f_6, f_8\}$ are associated with the horizontal directions, and $f_1$ and $f_5$ are purely horizontal. We can then build the propagation in such a way that we first act with a vertical shift on the first group, and then with a horizontal shift on the second group. A simple way to do this is to use a flag qubit to mark the substates which are to be shifted: regrouping operators are used to shuffle the substates to match the flag qubit appropriately and in correct order. The extension of this D2Q9 propagation to an equivalent three-dimensional D3Q27 model is straightforward: all we need to do is add another shift for the third dimension, and match the regrouping operators accordingly. The propagation routine for the D2Q9 model is outlined in Figure~\ref{fig:circuit-propagation-d2q9}.

Note that the scheme functions mostly in superposition for both time-levels $f_{\alpha}^k$ in the full OSSLBM model. The only necessary modification here is to make sure that the substates corresponding to the second time-level are shifted a \emph{double step} instead. If the shift routine is based on the quantum Fourier transform~\cite{shakeel2020} (QFT) as outlined in Figure~\ref{fig:circuit-propagation-d2q9}, this is easily done by simply adding a controlled phase gate block that amounts to doubling the phase translations on the relevant substates. Thus the price from moving from the Tau1 model to the full two-level model is very low. 

In terms of gate complexity, the regrouping operators are of constant size for a given lattice configuration and do not depend on the size of the lattice. These operators can be implemented with simple sequences of (possibly multi-controlled) $\cx$ gates. The main sources for gate complexity are the shift operators, which depend on the lattice size. In the example here we use the QFT-based shift, which has a quadratic gate complexity with respect to the number of lattice qubits due to the QFT routine. However, the number of phase shift gates is linear in terms of the number of lattice qubits. Thus, for each lattice dimension we need two QFT blocks (a QFT block and its inverse) each covering the number of qubits related to that dimension, and a set of phase shift gates depending linearly on the number of lattice qubits. The second time-level doubles number of phase shift gates, but there is no need for additional QFT blocks.

One can naturally replace the QFT-based shift with alternative methods, for example the parallel multi-controlled shift routine~\cite{shift-preprint}.  The cascade parallel shift is perhaps especially noteworthy in that the shift routine can be implemented with a linear complexity with the help of some ancilla qubits (the size of the ancilla register roughly the same as for the largest dimensional subset in the lattice register), that is, logarithmic with the number of lattice sites. The double-stepping for the full model is a simple extension also in this case: effectively one needs only to balance out the first decomposition block in the cascade shift for the substates marked for the second time level. This has a relatively low linear gate overhead over the single-stepping routine.

\subsection{Integration: back to macroscopic variables}

Reversing the collision step after the propagation, we now wish to move back from the local distribution functions to global macroscopic fields on the lattice. This way we complete the link between the local collision dynamics and the evolution of the global lattice state.

The equations implementing this mapping are simple (Eqs.~(\ref{eq:macros-oss-density}--\ref{eq:macros-oss-momentum})), stemming from the local integration of the relevant distribution functions over each lattice cell. On the discrete level we want to simply sum together the distribution functions for the relevant macroscopic fields $\rho$ and $\vu$.
Since the integration is local, we can devise operators that act in superposition over the whole lattice. In other words, the operators here act only on the register $s$, and possibly on some ancillae in the register $a$.

We make use of the simple observation concerning $\hgate$ gates on the level of amplitude coefficients to the basis states. Let us consider first the very simple 1D model D1Q2, where there are only two contributing distributions over each site: $f_1$ and $f_2$. The amplitude substates can be encoded over a single superposition qubit, and acting with an $\hgate$ gate on this qubit we get the amplitude map
\[
    (\alpha, \beta) \mapsto \frac{1}{\sqrt{2}} (\alpha + \beta, \alpha - \beta) , 
\]
where we interpret $\alpha$ and $\beta$ as amplitude arrays containing the data for $f_1$ and $f_2$ with respect to the lattice register. Setting aside the normalization coefficient for the time being, these amplitudes now contain \emph{exactly the data for the desired macroscopic density and momentum terms $\rho$ and $\vu$}.

The generalization to more higher-dimensionals models is rather straightforward. Specifically, we can form the field components for the D2Q9 model with just controlled $\hgate$ gate operators. In fact, two controlled $\hgate$ gates are enough to retrieve all the relevant terms, although in a decomposed form. If we first take a subset of the register $s$ representing the distributions $f_k$ paired as follows:
\[
    (f_1, f_5), (f_2,f_6), (f_3, f_7), (f_4, f_8) ,
\]
and then act with a controlled $\hgate$ gate targeted at each pair, we get distinguishable amplitude arrays carrying
\[
    (f_1 + f_5, f_1 - f_5), (f_2 + f_6, f_2 - f_6), (f_3 + f_7, f_3 - f_7), (f_4 + f_8, f_4 - f_8) .
\]
Further rearrangement gives us subsets
\begin{align*}
    & (f_1 + f_5, f_2 + f_6, f_3 + f_7, f_4 + f_8) \\
    & (f_1 - f_5), (f_3 - f_7) \\
    & (f_2 - f_6, f_4 - f_8) ,
\end{align*}
which already betrays the structure required: In the first subset we have all the components for the density $\rho$ apart from the distribution $f_0$, which has been kept separate. The second subset yields partial components to the momentum $\vu$, and can be made complete by acting with a controlled $\hgate$ gate on the pair in the last subset. Thus all in all we can gather the following subsets
\begin{align}
    &\rho: (f_0, f_1 + f_5, f_2 + f_6, f_3 + f_7, f_4 + f_8) \\
    &u_1: (f_1 - f_5, f_2 - f_4 - f_6 + f_8) \\ 
    &u_2: (f_3 - f_7, f_2 + f_4 - f_6 - f_8) ,
\end{align}
where each element is distinguishable as a substate in the register $s$.

In principle, this is enough data -- there is no need to combine these subsets into final fields. Boundary conditions can be just as well be applied on these components in superposition, and the collision at the next iteration can be likewise amended. The benefit from this is that there is no dissipation of the amplitudes that would come from further reduction. However, one needs to take care of the normalization coefficients that accompany each $\hgate$ gate. For example, the momentum terms $u_k$ have the weight $1/\sqrt{2}$ on the first component, and  $1/2$ on the second.

\begin{figure}[!htb]
    \centering
    \input{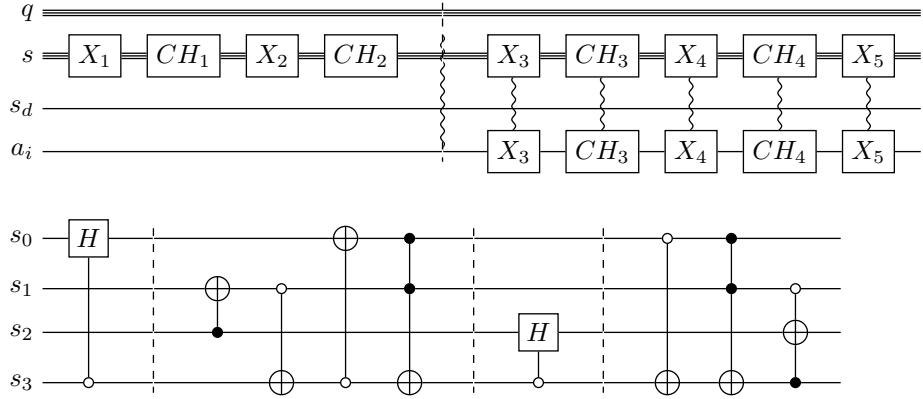}
    \caption{A circuit sketch for the D2Q9/Q17 integration. The barrier in the top figure separates the partial integration section that preserves amplitudes, and the completion to the full integration at the cost of amplitude dissipation. An ancilla qubit $a_i$ is used to aid in constructing the correct substate pairs for controlled $\hgate$ operations, and can be ultimately used to flag the wrong substates stemming from the full integration. The substate regrouping operators $X_k$ constitute of sequeces of multi-controlled $\x$ gates. The bottom circuit figure shows a possible implementation for the Tau1 (D2Q9) partial integration section in more detail, with the assumption that the substates in the register $s$ are ordered as $(f_1, f_5, f_2, f_6, f_3, f_7, f_4, f_8, 0_4, f_0, 0_3)$, where $0_n$ indicates $n$ zero amplitude blocks.}
    \label{fig:circuit-integration}
\end{figure}

Nevertheless, it may be required to complete the sums, for example to match some measurement operation. Additional $\hgate$ gates and possibly some rescaling rotations can be used to this end. An example implementation is sketched in Figure~\ref{fig:circuit-integration}, which can be applied for both the D2Q9 and the D2Q17 models, along with a particular circuit implementation for the D2Q9 model.

We use regrouping operators $X_k$ to arrange the distribution terms suitably for the sums. As with the regrouping operators in the propagation, we can employ sequences of multi-controlled $\x$ gates over the register $s$ (and, possibly, over a small set of ancilla qubits to make the substate management easier). Additional diffusion is unavoidable in this case, as the fields now carry $1/4$ of the desired amplitudes (pre-emptive scaling notwithstanding).

The full model extension from the Tau1 circuit is straightforward. The local sums affect both timelevels in superposition as indicated by the non-active qubit $s_d$. The major difference between the Tau1 and the full model is that in the latter we need to manage the extra substates holding the initial data. This may lead to some additional controls and slightly more complex blocks for the operators $X_k$.

As the local integration step is effectively a reversal of the collision operator, it does not depend on the size of the lattice. Thus for a given lattice configuration, its computational complexity can be considered as constant. The gate complexity does depend on the number $m$ of discrete velocities, as for each additional velocity there will be more terms in the sums of Eqs.~(\ref{eq:macros-oss-density}--\ref{eq:macros-oss-momentum}). This dependence is linear, with a complexity of $\bigO(\log^2(m))$ for the regrouping operators. In the D2Q9 case as in Section~\ref{sec:example_airfoil}, the number of $\cx$ gates required for ideal (all-to-all connectivity) circuit is around $50$.

\subsection{Enforcing boundary conditions}\label{sec:boundary_conditions}

Boundary conditions can be divided into two groups: external boundaries spanning the computational domain and inner boundaries outlining an object/obstacle within the domain. In this section we will deal with the domain boundaries.

The boundary conditions in the OSSLBM are implemented using the macroscopic field variables (see Section~\ref{sec:intro-boundary}), in contrast to the standard LBM where one needs to deal with the distributions directly. This means that the domain boundary conditions can be enforced with the standard Dirichlet and Neumann conditions for the density and momentum fields. 
However, the two time levels necessitate a specific treatment for two layers of boundary cells~\cite{qin2022}. 
Furthermore, the propagation subcircuit carries an inherent periodic condition affecting the two layers, and this needs to be amended for.
We deal with this by essentially ignoring the amplitude data of the second time-level on these lattice sites, and apply the boundary conditions proper on the first time-level only.

To this end, we need to modify the amplitudes at the domain boundary sections of the lattice. This process has three parts: locating and shifting the relevant amplitude data on the lattice to a zero amplitude subspace, modifying it according to the prescribed conditions, and finally transferring back the modified data to the original location.

\begin{figure}[!htb]
	\centering
\newcommand{\boundaryFigWidth}{0.7\textwidth}
\newcommand{\gridStep}{0.75}
\newcommand{\gridSize}{5.25}   
\newcommand{\gridNx}{7}
\newcommand{\gridNy}{7}
\newcommand{\cellSepX}{3.2}
\newcommand{\cellSepY}{2.2}
\tikzset{
    boundary arrow shorten/.initial=14pt,  
    boundary arrow/.style={
        -latex, thick,
        shorten >={\pgfkeysvalueof{/tikz/boundary arrow shorten}},
        shorten <={\pgfkeysvalueof{/tikz/boundary arrow shorten}},
    },
    boundary dot/.style={circle, fill=black, inner sep=1pt},
    boundary dot radius/.initial=1pt,   
    boundary dot label/.style={right, font=\small, inner sep=0.5pt},
    boundary grid/.style={black, very thin, opacity=0.25},
    boundary axis/.style={->},
    boundary region y/.style={draw=boundaryColorY, fill=boundaryColorY!15, opacity=0.7},
    boundary region x/.style={draw=boundaryColorX, fill=boundaryColorX!15, opacity=0.7},
    boundary center label/.style={font=\large},
    boundary operator arrow/.style={-latex, thin, bend left=12},  
}
\colorlet{boundaryColorY}{CBred}    
\colorlet{boundaryColorX}{CBcyan}  
\newcommand{\boundaryDotLabelSep}{0.12}  
\newcommand{\boundaryDotLabel}[2]{\node[boundary dot label, anchor=west] at ([shift={(\boundaryDotLabelSep,\boundaryDotLabelSep)}]#1) {#2}}
\newcommand{\boundaryDotAt}[1]{\fill[black] (#1) circle (\pgfkeysvalueof{/tikz/boundary dot radius});}

\begin{adjustbox}{width=\boundaryFigWidth}
\begin{tikzpicture}[
    x=\gridStep cm,
    y=\gridStep cm,
    every node/.append style={inner sep=1pt},
]

\newcommand{\drawGridOnly}[1]{%
    \draw[step=\gridStep, boundary grid] (0,0) grid (\gridSize,\gridSize);
    #1
}
\newcommand{\drawAxesOnly}{%
    \draw[boundary axis] (0,0) -- (1,0) node[right] {$x$};
    \draw[boundary axis] (0,0) -- (0,1) node[above] {$y$};
}

\begin{scope}[local bounding box=gridTL]
    \drawGridOnly{
        \draw[boundary region y] (-0.5,-0.25) rectangle (1.125,5.5);
        \draw[boundary region y] (4.125,-0.25) rectangle (5.75,5.5);
        \draw[boundary region x] (1.125,-0.25) rectangle (4.125,1.25);
        \draw[boundary region x] (1.125,4.25) rectangle (4.125,5.5);
        \coordinate (n00) at (0,0);
        \boundaryDotAt{0,0};
        \boundaryDotLabel{n00}{$a_{0,0}$};
        \coordinate (n11) at (.75,.75);
        \boundaryDotAt{.75,.75};
        \boundaryDotLabel{n11}{$a_{1,1}$};
        \coordinate (nij) at (2.25,2.25);
        \boundaryDotAt{2.25,2.25};
        \boundaryDotLabel{nij}{$a_{i,j}$};
        \coordinate (nnm) at (4.5,4.5);
        \boundaryDotAt{4.5,4.5};
        \boundaryDotLabel{nnm}{$a_{N_x-2,N_y-2}$};
        \coordinate (nNN) at (\gridSize,\gridSize);
        \boundaryDotAt{\gridSize,\gridSize};
        \boundaryDotLabel{nNN}{$a_{N_x-1,N_y-1}$};
    }
\end{scope}

\begin{scope}[shift={(\gridSize + \cellSepX, 0)}, local bounding box=gridTR]
    \drawGridOnly{
        \draw[boundary region y] (-0.5,-0.25) rectangle (1.125,5.5);
        \draw[boundary region y] (4.125,-0.25) rectangle (5.75,5.5);
        \node[boundary center label] (opY) at (2.5,5.75) {$B_{y}$};
        \draw[boundary operator arrow] (opY) to[out=-45, in=180] (0.3,2.5);
        \draw[boundary operator arrow] (opY) to[out=45, in=180] (4.95,2.5);
    }
\end{scope}

\begin{scope}[shift={(0, -(\gridSize + \cellSepY))}, local bounding box=gridBL]
    \drawGridOnly{
        \draw[boundary region x] (1.125,-0.25) rectangle (4.125,1.25);
        \draw[boundary region x] (1.125,4.25) rectangle (4.125,5.5);
        \node[boundary center label] (opX) at (-.75,2.5) {$B_{x}$};
        \draw[boundary operator arrow] (opX) to[bend left=12] (2.5,4.9);
        \draw[boundary operator arrow] (opX) to[bend right=12] (2.5,0.5);
    }
\end{scope}

\begin{scope}[
    shift={(\gridSize + \cellSepX, -(\gridSize + \cellSepY))},
    local bounding box=gridBR
]
    \drawGridOnly{
        \draw[boundary region y] (-0.5,-0.25) rectangle (1.125,5.5);
        \draw[boundary region y] (4.125,-0.25) rectangle (5.75,5.5);
        \draw[boundary region x] (1.125,-0.25) rectangle (4.125,1.25);
        \draw[boundary region x] (1.125,4.25) rectangle (4.125,5.5);
    }
\end{scope}

\begin{scope}[shift={(-1, -(\gridSize + \cellSepY + 1))}]
    \drawAxesOnly
\end{scope}

\draw[boundary arrow] (\gridSize, \gridSize/2) -- (\gridSize + \cellSepX, \gridSize/2);
\draw[boundary arrow] (\gridSize, -\cellSepY - \gridSize/2) -- (\gridSize + \cellSepX, -\cellSepY - \gridSize/2);
\draw[boundary arrow] (\gridSize/2, 0) -- (\gridSize/2, -\cellSepY);
\draw[boundary arrow] (\gridSize + \cellSepX + \gridSize/2, 0) -- (\gridSize + \cellSepX + \gridSize/2, -\cellSepY);

\end{tikzpicture}
\end{adjustbox}
	\caption{Mapping the boundary data of a 2D array $a$ on a lattice grid of size $N_x \times N_y$. First we locate the lattice sites matching with the two-cell layer of the boundary (upper left). The amplitude data on these layers is then moved into a zero subspace. The operators $B_{y}$ and $B_{x}$ (upper right and lower left) are used to enforce the boundary conditions on the amplitude data. Finally, the amplitudes are transferred back to the original position (lower right). Extending to 3D is straightforward.}
	\label{fig:boundary}
\end{figure}
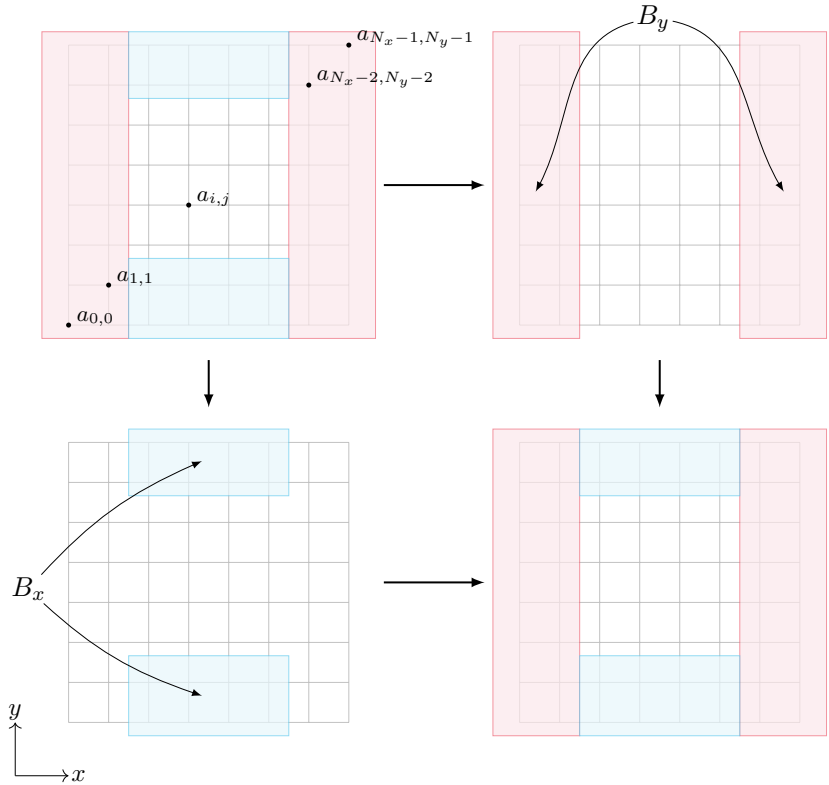
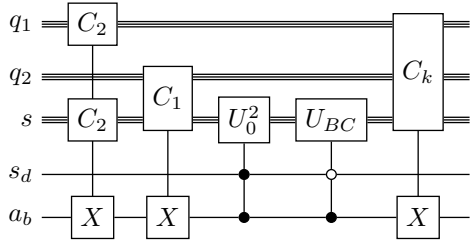
\begin{figure}[htb]
	\centering





      

\begin{tikzpicture}
    \begin{yquant}[operators/every barrier/.append style={red, thick}]
        qubits {$q_1$} q1;
        qubits {$q_2$} q2;

        qubits {$s$} s;
        qubit {$s_d$} sd;
        qubit {$a_b$} a; 

        [name=c2] box {\Ifnum\idx<2 $C_{2}$\Else $X$\Fi} (q1), (s), (a);
        \draw (c2-0) -- (c2-1) -- (c2-2);

        [name=c1] box {\Ifnum\idx<1 $C_{1}$\Else $X$\Fi} (q2,s), (a);
        \draw (c1-0) -- (c1-1);

        [name=b0] box {$U_0^2$} s | a, sd;

        [name=bc] box {$U_{BC}$} s | a, ~sd;
        
        [name=rev] box {\Ifnum\idx<1 $C_{k}$\Else $X$\Fi}  (q1,q2,s), (a);
        \draw (rev-0) -- (rev-1);
      
    \end{yquant}
\end{tikzpicture}
	\caption{A quantum circuit sketch for applying a uniform domain boundary condition in 2D models. The blocks $C_{k}$ refer to controls matching with the boundary parts of the lattice dimension $k$ for the two-cell layers, for given field amplitudes over register $s$. The gate $U_0^2$ zeroes the second time-level amplitudes on this boundary layer, and the gate $U_{BC}$ is the circuit implementation of the condition itself on the boundary amplitudes of the first time-level. As the last step, the modified amplitudes are shifted back to the substate with $a_b = 0$. Here we assume that a single ancilla qubit $a_b$ is used to modify these amplitudes; concrete implementations may use more. Similarly, some gate operations could be combined and, depending on the boundary condition, the order of the gate sequence may differ slightly.}
	\label{fig:circuit-boundary}
\end{figure}

First, locating the relevant data on the lattice splits into choosing the lattice sites matching with the boundary, and choosing the field variable to which the condition is applied. Due to the two-time level structure, we need to pick the field data at the two-cell layers along each solid boundary (upper left part in Figure~\ref{fig:boundary}). 
First we transfer the data matching the boundary cells into a zero-amplitude subspace to make the manipulation easier. This can be done with the help of an ancilla register and using a sequence of operators $C_kX$, each containing multi-controlled $\x$ gates corresponding to boundary lines along lattice dimension $k$, and operating on a subset of the lattice register $q$ according to the chosen spatial encoding (see Figure~\ref{fig:circuit-boundary}). 

As the second step, we apply the boundary operators $B_k$ on the transferred data, index $k$ depending on the dimension. What these operators exactly do depends on the type of the boundary condition. For example, in the 2D case such as in Figure~\ref{fig:boundary}, two columns encompass two vertical boundary layers (the red blocks), and we apply the operator $B_y$ on the amplitude subset $a_y(i,j)$ matching these cells. For a zero-gradient boundary condition, this would be the following:
\begin{equation}
    B_y:    
        \begin{cases} 
            a_y(0,j) = a_y(1,j) \\
            a_y(N_x-1, j) = a_y(N_x-2,j) ,
        \end{cases}
\end{equation}
that is, the data on the inner layers of the boundary cells is replicated at the outer layers. Similarly, for a zero-Dirichlet condition we would switch the data on the outer layers with zeros. Similar treatment then follows with the operator $B_x$ on the horizontal boundary layers.

Finally, we transfer the updated macroscopic data on the two boundary lines back to the original location followed by the summation of the final macroscopic variables (Eqs.~(\ref{eq:macros-oss-density},~\ref{eq:macros-oss-momentum})), corresponding to the lower right part in Figure~\ref{fig:boundary}. For this, we use the same set of operators $C_kX$ as in the first step, possibly with some additional multi-controlled $\x$ gates to match the time-levels correctly.

\subsubsection{Combining the two time-levels}

The final summation of the time-levels according to Eqs.~(\ref{eq:macros-oss-density},~\ref{eq:macros-oss-momentum}) can in principle be done with a controlled $\hgate$ gate. This then gives us the final state, encoding the evolved macroscopic field data matching with the boundary conditions at the domain boundaries.

However, there are two details that need to be addressed. First is that the boundary treatment above should properly apply on the first time-level only; for the second time-level, we need to make sure that the boundary data is not mixing with the conditions applied. This is relatively simple as it can be done by choosing the gate controls carefully, and by making sure that the second time-level boundary data is zeroed.

The other detail concerns the weights $c_k$ for each time-level $k$:
\begin{equation}
   c_1 = \frac{3 - 2\tau}{2 - \tau} , \quad  c_2 = \frac{\tau -1}{2 - \tau} ,
\label{eq:timelevel-coeffs}
\end{equation}
applied as in Eqs.~(\ref{eq:macros-oss-density},~\ref{eq:macros-oss-momentum}). Since the amplitude data is always subject to the global normalization, and there is no simple unitary way to scale the amplitudes, we need to make sure that the weights are applied relative to the rest of the field data. We have two (non-exclusive) ways to deal with this: using additional scaling terms when constructing the coefficients in the collision matrix, and re-scaling the amplitude data with controlled rotation gates in the boundary circuit. Note that such a rotational \quotes{scaling} will lead to some dissipation of amplitudes.

There is no single scheme that could be applied to all cases, as the possible rotations depend on the parameter $\tau$: for example, it is be impossible to devise a single rotation that scales amplitudes with a coefficient $c_k > 1$. In such cases, one must consider a relative scaling using coefficients like $c_j/c_k$.

\subsection{Immersed objects}\label{sec:immersed_object}

To simulate the effect of immersed objects in the domain, we can apply a zero-fluctuations condition on those lattice sites as prescribed by the masking method introduced in Section~\ref{sec:intro-maskingmethod}. 
However, due to the two-time level structure, only the two cell layers near the object boundary are affected by the computation. It is therefore sufficient to impose the zero condition only to those two lines; this may be beneficial for reducing the gate complexity.

We introduce multi-controlled $\x$ gates operating on the sites along the two layers of boundary cells inside the object subdomain (first being the outer boundary layer and the second a layer of cells inside), targeted at an ancilla qubit. This transfers zero amplitudes from a zero subspace into to the object location in a similar way to the solid boundaries.

We are left with some accumulation of residual amplitudes behind the ancilla, leading to some diffusion of the field data in the sense of probabilities. How severe this is depends on the use case -- typically, the size of the object boundaries tend to be small in comparison to the full domain.

\section{The scope and complexity of the quantum LBM}%
\label{sec:analysis}

\subsection{Linear and nonlinear models}

As discussed in Section~\ref{sec:lbm}, the LBM can be used for modeling of various physical process. Depending on the physics being simulated, these processes can be divided in two major groups: linear and nonlinear dynamics. 
Typically, nonlinear problems are linearized to enable straightforward computation and use of well-established linear techniques. On the other hand, the linear foundation of quantum mechanics imposes a significant limitation in terms of how efficiently we can simulate complex dynamics on quantum computers. 

In the LBM models, there are two main sources for nonlinearity. The clearest case is when the derivation of the equilibrium distribution functions from the macroscopic field variables (Eq.~\eqref{eq:definition-distributions-from-macros}) is nonlinear. On the other hand, the governing discrete equation Eq.~\eqref{eq:discrete-boltzmann} can also have nonlinearities, for example if it is affected by ambient materials with nonlinear properties (for example, magnetic materials in electrodynamics). Here we focus on the first class.

In the context of the presented quantum algorithm, the linear cases are easily implemented by first defining the distribution functions $\{f_{\alpha}\}$ to match the problem. Since the problem is linear, the resulting collision matrix $C$ can be used directly and turned into a quantum operator as in Section~\ref{sec:quantumlbm}.

The linearization of the equilibrium distribution equations in Eq.~\eqref{eq:definition-distributions-from-macros} is a common approach in posing nonlinear problems with quantum algorithms for the LBM~\cite{sanavio2024carleman, sanavio2024three,sanavio2025carleman,bastidazamora2026}. For example, by using the Carleman technique, the nonlinear mapping is first linearized into an infinite set of linear equations which are then truncated to a particular order and solved using the linear LBM model. 
Furthermore, the time-concatenation (see the discussion below) is here directly embedded into the process, therefore excluding the need for some specific time-stepping. However, despite addressing the nonlinearity, issues like truncation order, dimensionality of the collision operator, and low probability of the final solution makes this method currently implausible for practical applications. Quantum machine learning has been explored as well to include the nonlinear component of the equilibrium distribution functions~\cite{lacatus2025surrogatequantumcircuitdesign,Itani2025QMLLBM}. Nonetheless, current research lacks decisive evidence on the capabilities of these models to solve the nonlinear part for cases where subtracting these higher terms in the equations for equilibrium distributions results in large relative errors.

A hybrid quantum-classical method can be considered as an alternative way for solving nonlinear problems. The local structure of the linear collision operator is retained here, but with the inclusion of a larger number of macroscopic field variables. Particularly, instead of encoding just a set of distinct macroscopic variables, one includes also relevant product terms. This inevitably imposes a constraint in form of data extraction and re-initialization of the solution from the quantum system after each time step, which limits the efficiency of the algorithm. Nevertheless, we use this technique in Section~\ref{sec:example_airfoil} for an example implementation of a nonlinear Navier-Stokes model.

\subsection{Algorithmic complexity}

Assuming a fixed lattice configuration (for example, D2Q17), the gate complexity splits to roughly three parts: constant subroutines (collision and integration), purely lattice-dependent subroutines (propagation, domain boundaries), and lattice- and physics-dependent subroutines (embedded objects). By constant subroutines we mean that the operators are of constant dimension irrespective of the number of sites or complications in the physics model. On the other hand, the complexity of the propagation depends purely on the number of lattice sites, being either quadratic (QFT-based shift) or linear (cascade shift with ancillas) with respect to the number of lattice qubits. This is also dominating the complexity for the domain boundary, although the actual boundary conditions may change things slightly. Lastly, object embedding will depend mostly on the shape of the embedding, indirectly affected by the lattice size.

Leaving aside potential complications from complicated domain and object boundaries, it is worthwhile to note that the \emph{core algorithm} -- collision, propagation, and integration -- can be considered very efficient. The number of gates required is driven by the propagation, with $\bigO\left(\log^2_2(N)\right)$ gate-complexity as the worst case. This requires $\log_2(N) + c$ qubits, where $c$ is a relatively small constant factor depending on the lattice configuration and choices in circuit optimization; for example, in Section~\ref{sec:example_airfoil} we build a D2Q9 model with $c=8$.

All of the above sidesteps state preparation and eventual measurement. We will briefly discuss these aspects, too, but remark that this is a general problem for any data-intensive quantum algorithm. However, there are good reasons to believe that one can leverage the features of CFD to ease both of these pain points, as in many applications it may not be necessary to prepare or readout very precise amplitude data. 

\subsection{State preparation and measurements}

For a general quantum state where every lattice site and every field has a different amplitude, the computational complexity of the state encoding is $\bigO(N)$, with $N$ the number of lattice sites.
The computational complexity of this step is highly dependent on the entanglement of the initial state. However, for a modestly entangled initial state, we can leverage matrix product state based methods for state preparation. These methods scale for very large systems with tensor cross interpolation. 

In \cite{kerppo2025} the entanglement entropy of the target quantum state is minimized variationally. Depending on the lattice size, we can either represent the amplitudes exactly or use tensor cross interpolation to load the coefficients according to some analytical distribution. This involves a manageable classical training overhead, which depends on the exact representation of the state and the depth of the variational circuit.

We emphasize that the initial state encoding is an area of active research. However, there are already methods with a reasonable scaling that can be used to prepare relatively simple initial states. It is natural for many practical use cases to build on top of such states, allowing either the initial encoding of more complex initial conditions or more efficient state preparation circuits compared to existing methods. In particular, we can often assume extra structure -- such as symmetries or smoothness -- from the states encoding CFD-relevant data, and do not need to account for full generality.

Similarly, performing full quantum state tomography at the end of the circuit execution is unfeasible even for a modest number of qubits. 
Rather than striving for full field data, the question hinges on extracting meaningful global information at the end of the full simulation loop. For example, in linear acoustics a natural choice is the acoustic energy which is a quadratic function of the lattice fields $\rho$ and $\vu$. The acoustic energy corresponds to a diagonal observable, which can be measured with a single measurement setting. In particular, estimating the acoustic energy only requires a computational basis measurement of the qubits in the $s$ register. We analyze this measurement step in more detail in Section~\ref{sec:example_acoustics}.

In Section~\ref{sec:example_airfoil}, we exploit several simplifications for both the state preparation and the state tomography (Appendix~\ref{sec:appendix-tomography}) in order to simulate a small-scale nonlinear problem. While not everything used there can be scaled to real-life scenarios, these methods nevertheless point to the possibility of tailoring generic setups for further advantage in CFD.

\subsection{Connecting time steps}

As the LBM involves a discretization of time-dependent equations, it needs to include a time-stepping scheme. Advancing the computations in discrete time steps without performing a costly measurement-encoding operation after each time step is a necessary prerequisite for achieving any real quantum advantage. In general, there is an exponential cost for the measurement and re-encoding of the variables, especially since the measurement would require a partial tomography of the state to account for negative amplitudes.

Consequently, we need a suitable technique for connecting different time levels where the extraction of the information is needed only at the end of simulation. The choice of the method depends on the type of the problem being simulated (linear vs nonlinear).

\begin{figure}[htb]
	\centering
\begin{tikzpicture}
    \begin{yquant}

    qubits {$q$} q;
    qubits {$s$} s;
    qubits {$a$} a;
    qubit {$a_s$} as;

    box {$\operatorname{ENCODING}$} (q,s);
    box {$\operatorname{QLBM}_{t=1}$} (q,s,a,as);
    measure as;
    box {{$\operatorname{QLBM}_{t=2}$}} (q,s,a) ~ as;
    measure as;
    box {{$\operatorname{QLBM}_{t=3}$}} (q,s,a) ~ as;
    text {$\dots$} q, s, a, as;
    
    box {{$\operatorname{READOUT}$}} (q,s,a,as);    

    \end{yquant}
\end{tikzpicture}
	\caption{A dynamic circuit sketch for the time-stepping procedure. Here, the qubit $a_s$ is used as a flag qubit for time-stepping, and all the residual garbage is concentrated on states with $a_s = 1$. Thus, measuring $a_s = 0$, we can advance to the next step and repeat the circuits for the LBM.}
	\label{fig:dynamic-circuit}
\end{figure}
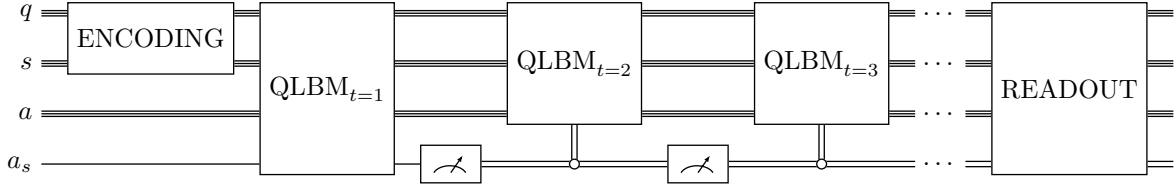

Again, for linear problems this is to some extent an easier task. Since we need information about macroscopic variables at the end of each time step, subsequent application of the quantum routines for the next step is in general conditioned by securing clean ancillas required by the non-unitary computations of the collision and integration steps.
To manage the cleanup after each time step efficiently, the most promising method seems to be mid-circuit measurements and dynamic circuits, such as in \cite{wawrzyniak2025dynamic}. By concentrating the the residual garbage behind a flag ancilla qubit, we can advance to the next time step by first measuring the flag qubit and then applying the next step conditionally on measuring $0$ (see Figure~\ref{fig:dynamic-circuit}, where $a_s$ acts as a flag qubit). 
In this way each new time step starts with a clean ancilla register $a$, which is sufficient for the continuation of the computation. Once the simulation is finished the information is extracted from the system by using the post-selection procedure.

Time concatenation for the nonlinear cases is mainly dependent by the type of the linearization being used. 
In the case of the Carleman linearization, the colliding and propagating distribution functions (and other Carleman variables) are initially encoded across different time levels. The time-stepping is then naturally encoded into the process, and no specific treatment is required. However, the practical application of the Carleman linearization is still out of reach due to the issues mentioned previously.    
\section{Solving 2D linear acoustics}%
\label{sec:example_acoustics}

As a first example case we look into linear acoustics. In this section we consider a simple statevector-simulated model in comparison to an analytical solution, and then discuss in more detail how to extract meaningful information from measurements. Specifically, we look into a formulation of acoustic energy that allows us to devise a simple measurement scheme for the quantum algorithm.

\subsection{Evolution of a Gaussian pulse}

Linear acoustics is the study of sound waves of relatively small amplitude. This can be modeled using the basic principles of fluid mechanics, neglecting viscosity and thermal conductivity effects. We can build the LBM on the hydrodynamic equations (see Appendix~\ref{sec:appendix-lbm}):
\begin{align}
    \partial_t \rho + \nabla \cdot (\rho \vu) &= 0 \\
    \partial_t \rho \vu + \nabla \cdot ( \rho \vu \otimes \vu + p \bm{I}) &= \nu \nabla^2 \vu
    \label{eq:acoustic-basic} ,
\end{align}
where we deal with acoustic density $\rho$, acoustic particle velocity $\vu$, and acoustic pressure $p = c_s^2\rho$, with kinetic viscosity $\nu$ and characteristic lattice speed $c_s$. 

The model unknowns $(\rho, \vu)$ are defined as the \emph{fluctuation} components to the above fields (denoted here with the same symbols for simplicity). The distribution functions can be defined as in Eq.~\eqref{eq:acoustics}, and from this we can construct a linear matrix $C$ for the collision subroutine in the quantum circuit.

\begin{figure}[htbp]
    \centering
    \includegraphics[width=0.85\linewidth]{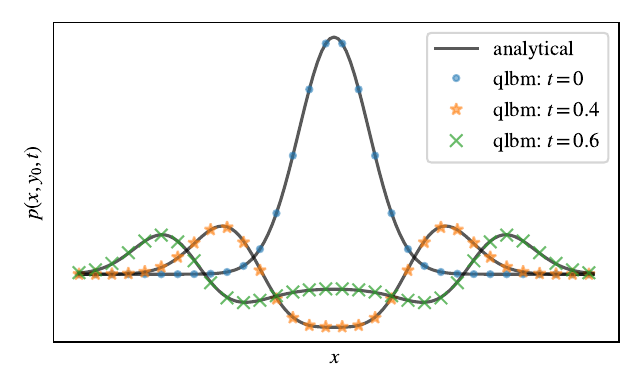}
    \caption{Evolution of the acoustic pressure fluctuation $p$ along the $x$-direction, with a qualitative comparison of the quantum simulation to an analytical solution. The initial condition corresponds to a Gaussian profile $p(x,y) = e^{-\beta ( x^2+y^2 )}$. We get the average relative error $\bar L_2^{rel} = 0.03$, which means a total discrepancy between the analytical and the quantum algorithm is below $3\%$ during the evolution for $t\in[0,0.8]$. }
    \label{fig:acoustic2d}
\end{figure}
As an example case, we use a well-known analytical solution for 2D acoustics: a Gaussian acoustic pulse evolution, shown in Figure~\ref{fig:acoustic2d} for a grid of $128\times 128$ points. The quantum model uses a total of 21 qubits, out of which 14 are for the lattice register $q$.  The results from a statevector simulation give an excellent match with the analytical solution.

\subsection{Energy measurement}

The last part of the algorithm after $T$ time steps is the measurement. Full state tomography scales as $\Omega(N)$ when the underlying quantum state contains $N$ non-zero amplitudes. This makes full state tomography unfeasible with few exceptions. However, we are still able to extract meaningful information without it. For multiphysics simulations we could attempt to extract energy, field averages or extreme values (such as velocity or pressure), gradients or other more easily obtainable data. We will explain how to obtain the acoustic energy with a simple measurement setup.

For linear acoustics, a natural choice is the acoustic energy which is a quadratic function of the macroscopic variables. The acoustic energy corresponds to a diagonal observable, which can be measured with a single measurement setting. In particular, estimating the acoustic energy only requires a computational basis measurement of the qubits in the $s$ register. 

Following \cite{rienstra1992introduction}, the acoustic energy can be expressed as
\begin{equation}
    E(S) = \frac{1}{2} \sum_{x \in S} {c_{phys}^2 \rho^2(x) + (u_1)^2(x) + (u_2)^2(x)} ,
    \label{eq:acoustic-energy}
\end{equation}
where $\rho(x)$ and $u_i (x)$ represent the fluctuating density and velocities over the lattice sites, $c^2_{phys}$ is a conversion factor relating to the speed of sound and  $x$ is summed over a subdomain $S$ of the lattice.

The full quantum state at the end of the simulation contains unclean ancilla qubits, so the state has to be cleaned up before we have access to the part that contains the macroscopic variables. This can be done by measuring each ancilla qubit to be in the zero state. In addition, there is a norm factor for the macroscopic variables that we have to keep track of. We will return to this after analyzing the measurement setup. For now, we assume we have access to the part of the quantum register containing the macroscopic variables and that the norm factor is known a priori.

We will denote with $\Hat{E}(S)$ the operator that corresponds to the acoustic energy over the domain $S$. This operator has two nonzero eigenvalues $c^2_{phys}/{2}$ and ${1}/{2}$ corresponding to the density and velocity parts of the register, so that $\langle \hat{E}(S) \rangle_{\psi} =  \langle \psi | \hat{E}(S) \psi \rangle$ equals the acoustic energy for a state $\psi$ corresponding to some encoding of the macroscopic variables. The variance of this observable is
\begin{equation}
    \Delta_{\psi}(\Hat{E}(S))^2 = \left( \frac{c^2_{phys}}{2} -  \langle \hat{E}(S) \rangle_{\psi}\right)^2 \Tr \left[ \rho\right] + \left( \frac{1}{2} -  \langle \hat{E}(S) \rangle_{\psi}\right)^2 \Tr \left[ \vu\right],
\end{equation}
where the terms $\Tr\left[ \rho\right]$ and $\Tr \left[ \vu\right]$ denote the probabilities to obtain a measurement outcome corresponding to the part of the register encoding density and velocities, respectively, over the subdomain $S$ when measuring the state $\psi$ in the computational basis. We note that when $S$ equals the whole domain the measurement simplifies to a computational basis measurement of the $s$ register, making the measurement particularly simple to implement. In this case we denote the corresponding observable as $\hat{E}$.

For a fixed input state  $\psi$ we can now estimate how many shots we need to estimate the acoustic energy. Since the variance is finite we can calculate the standard error as 
\begin{equation}
    \sigma_{\langle \hat{E} \rangle_{\psi}} = \frac{\sqrt{\Delta_{\psi}(\Hat{E})^2}}{\sqrt{n}},
\end{equation} 
where $n$ equals the number of samples. To obtain a simple estimate on the number of samples required to reach a specific standard error, we start by requiring that
\begin{equation}
    \sigma_{\langle \hat{E} \rangle_{\psi}} \leq \epsilon \langle \hat{E} \rangle_{\psi}  ,
\end{equation}
where $\epsilon$ denotes the desired level of relative accuracy. Solving for $n$, we obtain 
\begin{equation}
    n \geq \frac{\Delta_{\psi}(\Hat{E})^2}{\epsilon^2 \langle \hat{E}\rangle_{\psi}^2 }.
\end{equation}
To simplify further we note that, while the variance depends on the encoding state $\psi$, it is always upper bounded by the acoustic energy squared. We then obtain the simple upper bound for the number of samples $n \geq \epsilon^{-2}$ to reach a relative accuracy of $\epsilon$ on the acoustic energy. 

\begin{figure}[htbp]
    \centering
    \includegraphics[width=1.0\linewidth]{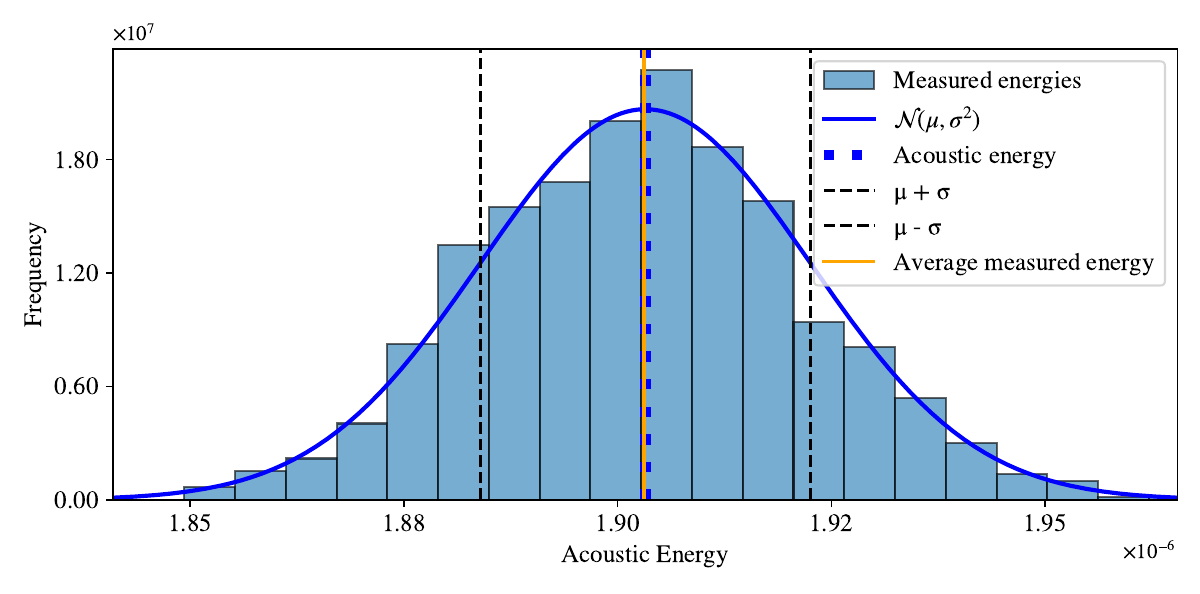} 
    \caption{A histogram for simulated acoustic energy measurements repeated 1000 times. }
    \label{fig:acoustic-energy-measurements}
\end{figure}

To confirm this scaling we simulated the acoustic energy measurements for 1000 times starting from a simple initial state and evolving it for a few time steps.
The exact acoustic energy was $\langle \hat{E} \rangle_{\psi} = 1.9 \times 10^{-6}$, while the variance was calculated to be  $\Delta_{\psi}(\Hat{E})^2 = 1.9 \times 10^{-8}$.
To reach a relative error of $\epsilon=0.01$, each individual experiment consisted of 3146 samples. As expected, the individual measurement outcomes followed a Gaussian distribution and the mean of all experiments very closely matched the exact energy. The measurement outcomes are illustrated in Fig.~\ref{fig:acoustic-energy-measurements}.

There are two additional factors that contribute overhead to the scaling of the acoustic energy measurement. First is related to the subdomain $S$ being measured. When $S$ is not the whole domain, we need to also measure the $q$ register to obtain information on the distribution of energy over the lattice sites. This is certainly feasible, but we will focus on the more simple case of measuring the whole domain.

The second factor comes from estimating the norm of the encoding state $\psi$, as generally this state is unnormalized initially. That is, the macroscopic values of $\rho$ and $u_i$ are encoded in a subnormalized state, but measuring the ancilla qubits has the effect of normalizing the state in each time step. The embedded non-unitary operations also cause the state to diffuse into the ancilla space which has to be taken into account. The magnitude of state diffusion depends on the initial state and the structure of the collision matrix. In practice the diffusion tends to stabilize after a few time steps. The normalization factor and state diffusion can be estimated by counting how many times the ancilla qubits are measured to be zero. The ratio of zero to non-zero ancilla counts then gives the required norm factor for the state directly. Rigorous accuracy bounds and confidence intervals can be derived with Hoeffding's inequality. Since the samples where the ancillas are measured to be non-zero are discarded from the acoustic energy measurement, the norm of the state will be estimated with a relatively high accuracy. Hence we do not analyze this factor further but omit it from further analysis.

\subsubsection{Example: dissipating energy}

\begin{figure}[htbp]
    \centering
    \input{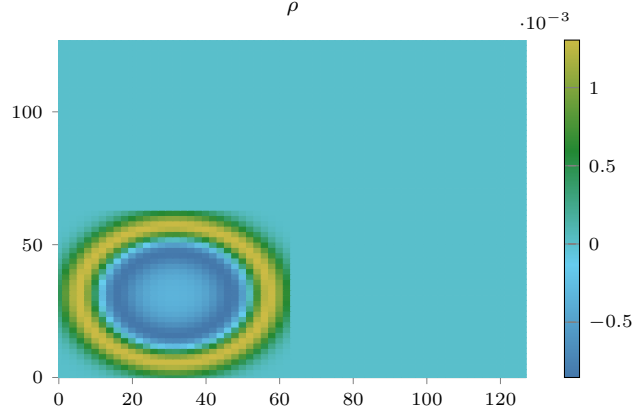}
    \caption{Initial state for the energy dissipation measurement. Here, a Gaussian pulse is placed at the bottom left quadrant of a $128\times128$ lattice, and the measurement is conducted over this $64\times64$ cell.}
    \label{fig:acoustic-initial}
\end{figure}

In this example we place a Gaussian pulse inside a $64\times 64$ region, which itself is inside a larger $128\times128$ lattice so that we can observe the energy dissipating outside the initial domain. This setup mimics open boundaries without needing to describe them through any explicit boundary conditions as such. The initial state is depicted in Fig.~\ref{fig:acoustic-initial}.

\begin{figure}[htbp]
    \centering

\begin{tikzpicture}
    \begin{axis}[
            width=0.95\linewidth,
            height=0.25\paperheight,
            axis x line=bottom,
            axis y line=left,
            axis line style={-},
            xlabel={Time-step},
            ylabel={Normalized energy},
            xmin=-1, xmax=39,
            ymin=0.45, ymax=1.15,
            legend pos=north east,
            grid=major,
            mark size=1.5pt,
            ]
            \addplot[thick, dashed, draw=CBblue, mark=*, mark options={fill=CBblue, solid}, mark size=2pt] table[x=timestep, y=theory]{acoustic_energy_figure.dat};
            \addlegendentry{Normalized acoustic energy}
            \addplot[thick, draw=CBred, mark=x, mark options={fill=CBred, solid}, mark size=3pt] table[x=timestep, y=measured_14]{acoustic_energy_figure.dat};
            \addlegendentry{Measured energy ($2^{14}$ total shots)}
            \addplot[thick, dotted, draw=CBgreen, mark=triangle, mark options={fill=CBgreen, solid}, mark size=3pt] table[x=timestep, y=measured_17]{acoustic_energy_figure.dat};
            \addlegendentry{Measured energy ($2^{17}$ total shots)}
    \end{axis}
\end{tikzpicture}
    \caption{The evolution of measured acoustic energy (normalized) for $2^{14}$ and $2^{17}$ total shots, against ideal statevector data. The effect of the shot count on the accuracy is well visible.}
    \label{fig:acoustic-energy-dissipation}
\end{figure}
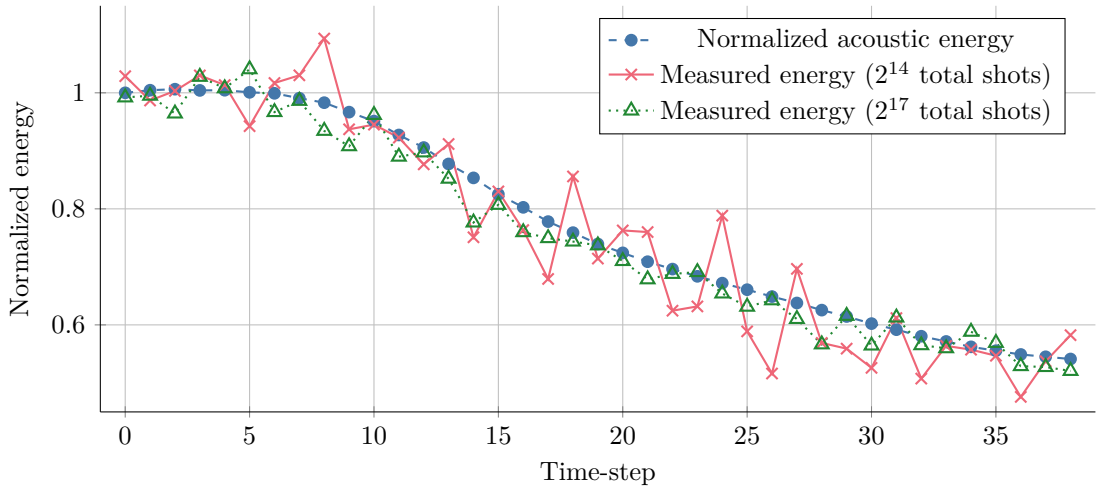

In Figure~\ref{fig:acoustic-energy-dissipation}, the results of the energy measurement are compared against known statevector data. Here, we use two different shot counts to visualize the effect of the shot noise in the measurement. As all computations are done in lattice units, we use normalized energy to provide qualitative analysis of the measurement -- the normalization is done with respect to the initial energy as given by the statevector data.
\section{Solving a 2D nonlinear airfoil problem on a quantum computer}%
\label{sec:example_airfoil}

In this section, we validate the practical performance of the OSSLBM quantum algorithm for a nonlinear problem describing a 2D flow around an immersed object. We execute the circuits on a noisy QPU, and compare the results to what is expected based on an ideal statevector simulation.

To this end, we define a simple 2D airfoil-like problem for the incompressible Navier-Stokes equations with a solid object inside the domain. The field evolution follows the equations
\begin{align}
    \partial_{x_i} \rho u_i &= 0 \\
    \partial_{t} u_i + \partial_{x_j}\left(u_iu_j + P\delta_{ij}\right) &= \nu \, \partial_{x_i, x_j}^2 u_i ,
    \label{eq:airfoil-nse}
\end{align}
where $\nu$ is the viscosity, $u_k$ the components of the velocity field (momentum $\vu$ from the abstract model), and $P = p/\rho_0$ is the normalized pressure, with $\rho_0$ a reference base flow value for the density $\rho = \rho_0 + \rho'$. We assume $\rho_0$ is uniform and steady, so that the equations can be formulated in terms of the density fluctuation $\rho'$. Dropping the apostrophe for notational simplicity, the pair $(\rho,u_k)$ then represents the unknown variables sought for in the computation.

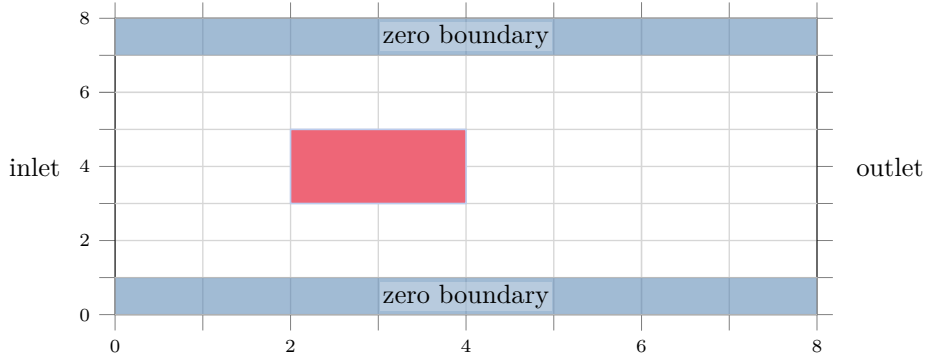
\begin{figure}[htbp]
    \centering

\pgfplotsset{
  schematicaxis/.style={
    width=0.7\linewidth,
    unit vector ratio=1 0.65,
    axis lines=box,
    axis line style={-},
    enlargelimits=false,
    xmin=0, xmax=8, ymin=0, ymax=8,
    xtick={0,2,4,6,8},
    ytick={0,2,4,6,8},
    extra x ticks={1,3,5,6,7},
    extra y ticks={1,3,5,6,7},
    extra x tick style={tick label style={opacity=0}},
    extra y tick style={tick label style={opacity=0}},
    major tick length=6pt,
    tick align=outside,
    grid=major,
    grid style={line width=.5pt, draw=CBgrey!60},
    tick label style={font=\scriptsize},
    xlabel={},
    ylabel={},
    clip=false,
  },
}

\begin{tikzpicture}[baseline]
  \begin{axis}[schematicaxis, name=schematic, xshift=0.15\linewidth]
    \addplot [draw=none] coordinates {(0,0) (8,8)};
    \fill [fill=CBred, draw=CBpaleblue, line width=0.6pt] (axis cs:2,3) rectangle (axis cs:4,5);
    \fill [fill=CBblue, fill opacity=0.5, draw=CBgrey, line width=0.4pt] (axis cs:0,0) rectangle (axis cs:8,1);
    \fill [fill=CBblue, fill opacity=0.5, draw=CBgrey, line width=0.4pt] (axis cs:0,7) rectangle (axis cs:8,8);
    \node [fill=white, fill opacity=0.25, text opacity=1, inner sep=1pt] at (axis cs:4,0.5) {zero boundary};
    \node [fill=white, fill opacity=0.25, text opacity=1, inner sep=1pt] at (axis cs:4,7.5) {zero boundary};
  \end{axis}
  \node [anchor=east] at ([xshift=-6mm]schematic.west) {inlet};
  \node [anchor=west] at ([xshift=4mm]schematic.east) {outlet};
\end{tikzpicture}
    \caption{Airfoil model setup.}
    \label{fig:airfoil-sketch}
\end{figure}

\begin{figure}[htbp]
    \centering
    \input{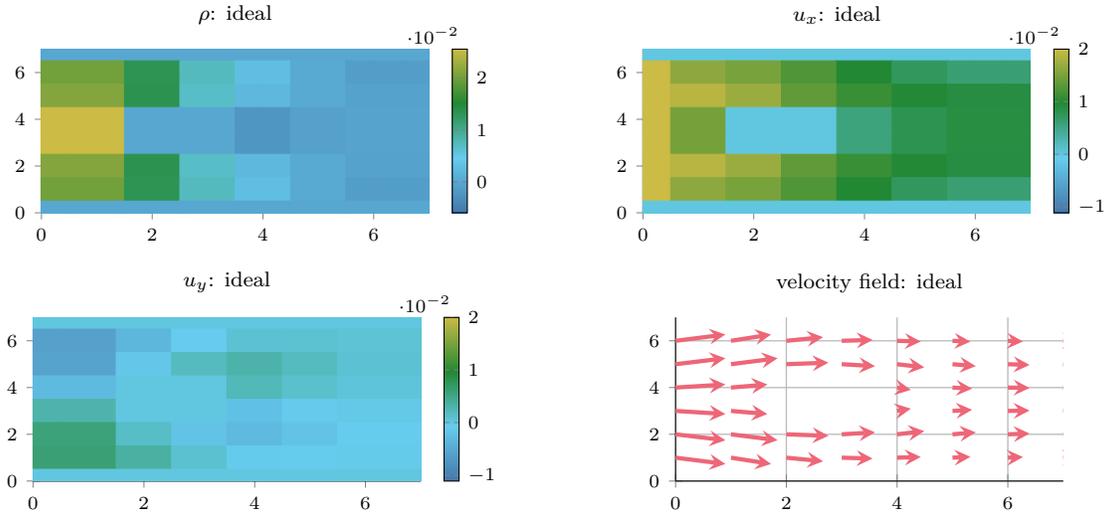}
    \caption{The results of ideal (statevector) simulation, and the reconstructed flow field after 15 time-steps. The simulation has not yet reached a steady state, but is already conforming to physically meaningful flow patterns.}
    \label{fig:airfoil-ideal}
\end{figure}

With the present OSSLBM model we can simulate these equations in the laminar regime. Here, we pose the problem on a $8\times8$ grid, with a $2\times2$ immersed square object inside, as in Figure~\ref{fig:airfoil-sketch}. 
We assume strict zero-boundary conditions at the bottom and the top of the grid, with a fixed inlet velocity prescribed at the left boundary layer. The inlet velocity is given in dimensionless lattice units as $u_x = 0.02, u_y = 0$ -- since the purpose of the experiment is to compare the ideal results to a noisy QPU simulation, we do not try to match the model with any physical data as such. In particular, the boundary conditions and the shape of object are chosen to minimize the circuit complexity. We also use the D2Q9 lattice configuration following the Tau1 simplifcation (giving a lattice viscosity of $1/6$). The state of this model after 15 time-steps is shown in Figure~\ref{fig:airfoil-ideal}, generated with an ideal statevector simulator from an initial state of uniform velocity matching with the inlet condition.

The associated quantum circuit requires six qubits to encode the lattice, and eight ancillary qubits for the problem definition. The three variables (and their nonlinear combinations) corresponding to the density and velocity fields are mapped to amplitudes indexed over the lattice register, each field occupying one substate in the superposition register $s$.

\subsection{Experiment setup}

The quantum-classical hybrid algorithm pipeline has the following structure for the example considered (see also Fig.~\ref{fig:overview-algorithm}): 
\begin{enumerate}
    \item The initial conditions for density and velocity (in the $x$- and $y$-directions) fields are defined together with the inlet velocity. Here we set the density and the $y$-velocity to $0$, and the $x$-velocity equal to the inlet velocity $0.02$ everywhere.

    \item\label{item:ibm-nonlinear} The array of non-linear terms \eqref{eq:nonlinear-terms} is prepared classically and loaded onto the quantum computer during state preparation. This requires three superposition qubits in the register $s$ to label the six non-linear terms.

    \item The quantum circuit implementing a single LBM step is executed on the hardware.

    \item Measurements with post-selection to account non-linear evolution in the QLBM are performed. From the collected shots tomography is performed (see below and Appendix~\ref{sec:appendix-tomography}) to restore the density and velocity fields.

    \item Each field is additionally renormalized, depending on the original norm of the non-linear array and the collision operator. Additionally the left boundary layer on the lattice is set to the inlet velocity $u_x = 0.02$.

    \item The process is repeated from Step~\ref{item:ibm-nonlinear} until solution convergence, or for a set amount of steps.
\end{enumerate}

To restore each field from measurements we introduce and perform function tomography, which represents the field as a finite sum of certain basis functions. Here we chose to use Chebyshev polynomials up to the 2nd degree for each coordinate. We also perform a grid transformation to map the immersed object to a single point and improve the tomography performance, which allowed to use polynomials of the lesser degree. For more details on the tomography method see the Appendix~\ref{sec:appendix-tomography}.

The norm of each field is restored from the corresponding observables. Indeed, the probability of measuring the bitstring which labels each field on the ancillary qubits is equal to the squared norm of the corresponding field. The computation of these observables requires only to trace out the lattice register from the full measurements.

The preparation of the non-linear terms in Eq.~\eqref{eq:nonlinear-terms} in Step~\ref{item:ibm-nonlinear} can be done directly, as classical computations can be done explicitly on the tomography-restored function. For large scale experiments, when the lattice register is large, it may also be necessary to load the array into an efficient intermediate representation such as matrix product state (MPS) before synthesizing the state preparation circuit. This can be done using tensor cross interpolation~\cite{fernandez2025tensorcrossinterpolation}, which can represent a given function as an MPS or as a tensor network, from which a quantum state can be efficiently prepared as shown in \cite{bohun2025entanglementscalingmatrixproduct,ballarin2025efficientquantumstatepreparation}.

\subsection{Circuit compilation and execution details}

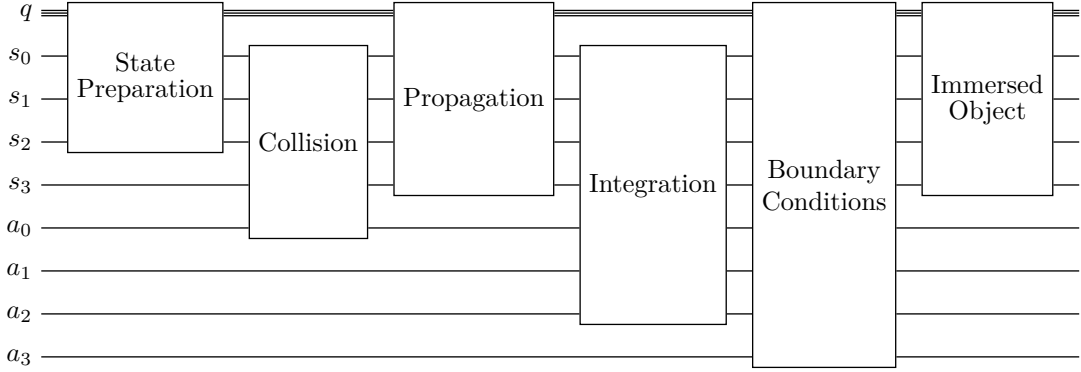
\begin{figure}[htb]
	\centering
\begin{tikzpicture}
    \begin{yquant}

    qubits {$q$} q;
    qubit s[4];
    qubit a[4];

    box {$\operatorname{\shortstack{State\\Preparation}}$} (q,s[0],s[1],s[2]);
    
    box {$\operatorname{Collision}$} (s,a[0]);

    box {$\operatorname{Propagation}$} (q,s);

    box {$\operatorname{Integration}$} (s,a[0],a[1],a[2]);

    box {$\operatorname{\shortstack{Boundary\\Conditions}}$} (q,s,a);

    box {$\operatorname{\shortstack{Immersed\\Object}}$} (q,s);

    \end{yquant}
\end{tikzpicture}
	\caption{The structure of the full LBM quantum circuit for the hardware experiment, showing the functional blocks and the affected registers. For more details on each step see Section~\ref{sec:quantumlbm}.}
	\label{fig:qlbm_final}
\end{figure}\label{fig:qlbm_whole}
We perform our experiment on the superconducting IBM Heron R3 device \code{ibm\_boston}, which has 156 qubits with a heavy-hex connectivity. The final  circuit structure is depicted in Figure~\ref{fig:qlbm_final} (see also Fig.~\ref{fig:overview-algorithm}). A direct and unoptimized compilation of the circuit results in a large depth, however. In order to increase its performance on the noisy hardware, we implemented a number of compilation improvements. 

First, note that the collision operator is prepared on four qubits of the $s$-register and is independent from the lattice size (see Sec.~\ref{sec:collision}). Instead of loading the input variables of Eq.~\eqref{eq:nonlinear-terms} directly, and implementing the collision step as the unitary operator, we absorb the collision operator into the state preparation. This can always be done classically, and results in performing the state preparation on two more qubits than originally. We tested MPS~\cite{bohun2025entanglementscalingmatrixproduct, kerppo2025} and more general tensor network~\cite{ballarin2025efficientquantumstatepreparation,Manabe_2025} techniques for state preparation and found that while the latter ones provide better performance for multi-dimensional data and multiple variables, tensor network based state preparation results in circuits with a long-range connectivity. After transpilation to a device with near linear topology, their performance advantage is typically lost due to additional swap gates introduced. In this experiment we thus used the simpler MPS state preparation in linear connectivity.

The second improvement concerns the boundaries. The tail of the circuit, which implements the boundary conditions (partially) and the immersed object (fully), consists only of $\x$, $\cx$ gates and their multi-controlled versions (see Sections~\ref{sec:boundary_conditions} and~\ref{sec:immersed_object}). These gates commute with measurements in the $\z$ basis, therefore could be backpropagated \cite{Fuller2026} and performed classically on the already measured bitstrings. The backpropagation greatly reduced circuit complexity, but puts a restriction on doing the measurements only in the computational basis. Therefore in this experiment we perform only absolute-value function tomography (see Appendix~\ref{sec:appendix-tomography}), when executing the circuit on the hardware. Since the problem of interest is a laminar flow with a symmetric object, we can restore the relative sign of the solution from its symmetries.

Lastly, we use optimized implementations of gates in the remainder of a circuit. For Toffoli and multi-controlled $\x$ gates we use relative-phase decompositions~\cite{maslov2016advantages} whenever possible. For multi-controlled rotation gates we use the efficient multi-controlled special-unitary implementation~\cite{vale2023decompositionmulticontrolledspecialunitary}.

The combination of these compilation steps allows us to reduce the $\cx$ gates count to around 300 gates in all-to-all connectivity, down from 800 for the direct implementation. The two-qubit gates count is reduced from 1825 to 540 after transpilation to the \code{ibm\_boston} quantum processor.

When executing the circuit on the hardware, additional error mitigation techniques were used. We use dynamical decoupling (DD) with XX sequences to account for large idle times in the circuit~\cite{Ezzell_2023}. We also perform Pauli Twirling  with 16 random instances~\cite{Wallman_2016} to symmetrize the noise effects. Additionally operator decoherence renormalization (ODR) is used to correct the observable expectation values, which define the norms of each field~\cite{Urbanek_2021}. 

In total 14 circuits were executed to perform the evolution from time step 1 to 15 iteratively, with a single additional ODR circuit, since the circuit structure does not change. For each circuit we performed $30,000$ shots. The shot number has to account for fact that the bitstrings must be post-selected, and only 3\% to 6\% of shots survive the post-selection (depending on the step) in the ideal simulation, and split to their corresponding fields. The actual tomography is performed only using a few hundreds of shots for density and the $x$-velocity fields, and few tens of shots for the $y$-velocity field.

\subsection{Experiment results}

\begin{figure}[htbp]
    \centering

\pgfplotsset{
  colormap={cbsequential}{[1pt] color(0)=(CBblue) color(1)=(CBcyan) color(2)=(CBgreen) color(3)=(CByellow)},
  colormap={cberror}{[1pt] color(0)=(CBdarkred) color(1)=(CBred) color(2)=(CByellow)},
  col1/.style={point meta min=-0.00598, point meta max=0.02555},
  col2/.style={point meta min=-0.01115, point meta max=0.02},
  col3/.style={point meta min=0, point meta max=0.01984},
  mainaxis/.style={
    width=\linewidth,
    height=0.65\linewidth,
    axis x line=bottom,
    axis y line=left,
    axis line style={-},
    every axis x line/.append style={-},
    every axis y line/.append style={-},
    enlargelimits=false,
    xmin=0, xmax=7, ymin=0, ymax=7,
    xtick={0,2,4,6},
    ytick={0,2,4,6},
    tick pos=left,
    major tick length=5pt,
    grid=major,
    grid style={line width=.5pt, draw=CBgrey},
    tick label style={font=\scriptsize},
    title style={font=\footnotesize},
    colorbar style={
      tick label style={font=\scriptsize},
      width=2mm,
      xshift=0mm,
    },
  },
  every colorbar/.append style={grid=none, major tick length=2pt},
}
\begin{subfigure}[t]{0.32\textwidth}
  \centering
  \begin{tikzpicture}
    \begin{axis}[title={$\rho$: ideal}, colorbar, colormap name=cbsequential, mesh/cols=8, col1, mainaxis]
      \addplot [matrix plot*, point meta=explicit] table [x=x,y=y,meta=value] {ibm_data/ibm_step15_density_ideal.dat};
    \end{axis}
  \end{tikzpicture}
\end{subfigure}
\begin{subfigure}[t]{0.32\textwidth}
  \centering
  \begin{tikzpicture}
    \begin{axis}[title={$\rho$: IBM}, colorbar, colormap name=cbsequential, mesh/cols=8, col2, mainaxis]
      \addplot [matrix plot*, point meta=explicit] table [x=x,y=y,meta=value] {ibm_data/ibm_step15_density_sim.dat};
    \end{axis}
  \end{tikzpicture}
\end{subfigure}
\begin{subfigure}[t]{0.32\textwidth}
  \centering
  \begin{tikzpicture}
    \begin{axis}[title={$\rho$: abs. error}, colorbar, colormap name=cberror, mesh/cols=8, col3, mainaxis]
      \addplot [matrix plot*, point meta=explicit] table [x=x,y=y,meta=value] {ibm_data/ibm_step15_density_abs.dat};
    \end{axis}
  \end{tikzpicture}
\end{subfigure}

\begin{subfigure}[t]{0.32\textwidth}
  \centering
  \begin{tikzpicture}
    \begin{axis}[title={$u_x$: ideal}, colorbar, colormap name=cbsequential, mesh/cols=8, col1, mainaxis]
      \addplot [matrix plot*, point meta=explicit] table [x=x,y=y,meta=value] {ibm_data/ibm_step15_x_vel_ideal.dat};
    \end{axis}
  \end{tikzpicture}
\end{subfigure}
\begin{subfigure}[t]{0.32\textwidth}
  \centering
  \begin{tikzpicture}
    \begin{axis}[title={$u_x$: IBM}, colorbar, colormap name=cbsequential, mesh/cols=8, col2, mainaxis]
      \addplot [matrix plot*, point meta=explicit] table [x=x,y=y,meta=value] {ibm_data/ibm_step15_x_vel_sim.dat};
    \end{axis}
  \end{tikzpicture}
\end{subfigure}
\begin{subfigure}[t]{0.32\textwidth}
  \centering
  \begin{tikzpicture}
    \begin{axis}[title={$u_x$: abs. error}, colorbar, colormap name=cberror, mesh/cols=8, col3, mainaxis]
      \addplot [matrix plot*, point meta=explicit] table [x=x,y=y,meta=value] {ibm_data/ibm_step15_x_vel_abs.dat};
    \end{axis}
  \end{tikzpicture}
\end{subfigure}

\begin{subfigure}[t]{0.32\textwidth}
  \centering
  \begin{tikzpicture}
    \begin{axis}[title={$u_y$: ideal}, colorbar, colormap name=cbsequential, mesh/cols=8, col1, mainaxis]
      \addplot [matrix plot*, point meta=explicit] table [x=x,y=y,meta=value] {ibm_data/ibm_step15_y_vel_ideal.dat};
    \end{axis}
  \end{tikzpicture}
\end{subfigure}
\begin{subfigure}[t]{0.32\textwidth}
  \centering
  \begin{tikzpicture}
    \begin{axis}[title={$u_y$: IBM}, colorbar, colormap name=cbsequential, mesh/cols=8, col2, mainaxis]
      \addplot [matrix plot*, point meta=explicit] table [x=x,y=y,meta=value] {ibm_data/ibm_step15_y_vel_sim.dat};
    \end{axis}
  \end{tikzpicture}
\end{subfigure}
\begin{subfigure}[t]{0.32\textwidth}
  \centering
  \begin{tikzpicture}
    \begin{axis}[title={$u_y$: abs. error}, colorbar, colormap name=cberror, mesh/cols=8, col3, mainaxis]
      \addplot [matrix plot*, point meta=explicit] table [x=x,y=y,meta=value] {ibm_data/ibm_step15_y_vel_abs.dat};
    \end{axis}
  \end{tikzpicture}
\end{subfigure}
    \caption{Ideal (state-vector) vs hardware executed (\code{ibm\_boston}) field data results after 15 time-steps, and the absolute error. Top: density $\rho$; middle: $x$-velocity $u_x$; bottom: $y$-velocity $u_y$. Each column shares the same color scale. The model assumes $\tau=1$, and a fixed inlet velocity at the left-hand side with $u_x = 0.02, u_y=0$ as dimensionless lattice units. The bottom and top boundaries carry the zero Dirichlet condition.}
    \label{fig:airfoil-ibm}
\end{figure}
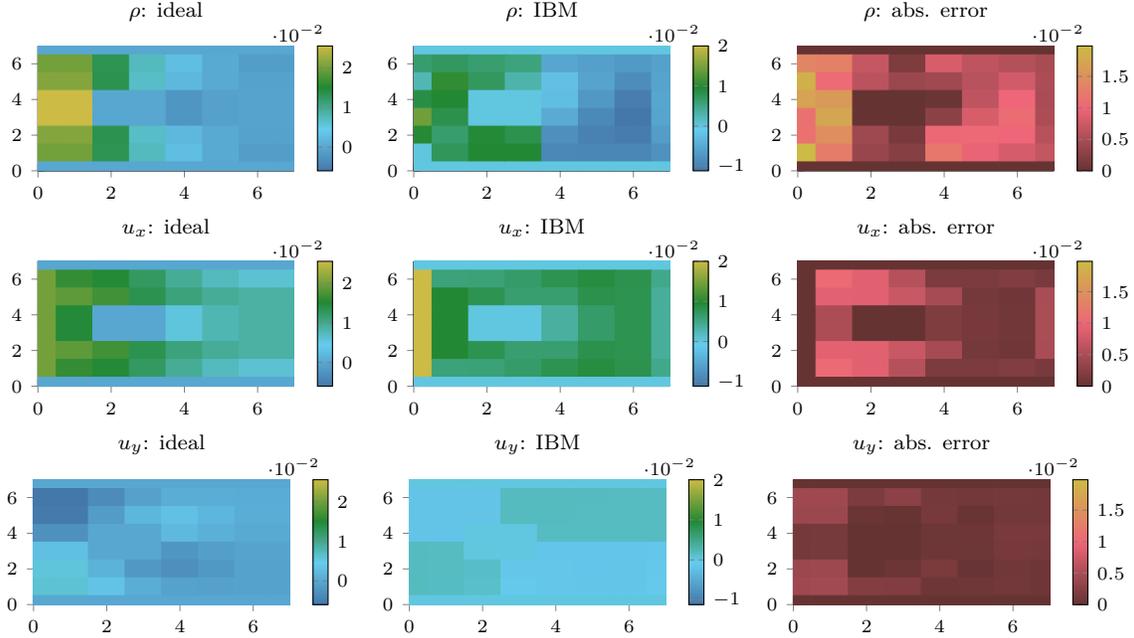
The results of the execution on the quantum computer are shown in Figure~\ref{fig:airfoil-ibm}, where we compare each field to the ideal data and show the absolute error at each lattice site. 
The obtained fields reproduce correctly the qualitative features of the desired state, though the $y$-velocity is somewhat \quotes{flattened} due to the quantum noise. The ideal solution is shown in full detail in Figure~\ref{fig:airfoil-ideal}. 

\begin{figure}[htbp]
    \centering


\begin{tikzpicture}
\begin{axis}[
    axis x line=bottom,
    axis y line=left,
    axis line style={-},
      width=0.95\linewidth,
      height=0.25\paperheight,
      xlabel={QLBM step},
      ylabel={MSE},
      xmin=0.5, xmax=15.9,
      xtick={1,4,...,15},
      ymin=0, ymax=0.00009,
      grid=major,
      grid style={line width=.2pt, draw=CBgrey!60},
      legend style={
            at={(0.98,0.98)},
            anchor=north east,
            legend cell align=left,
            font=\scriptsize,
      },
      every axis x label/.style={at={(ticklabel cs:0.5)}, anchor=near ticklabel},
      every axis y label/.style={at={(ticklabel cs:0.5)}, anchor=near ticklabel, rotate=90},
      tick label style={font=\scriptsize},
      label style={font=\small},
      axis line style={-},
      enlargelimits=false,
]
  \addplot [thick, dotted, color=CBblue, mark=*, mark size=2pt]
    table [x=step, y=statevector_err] {ibm_data/convergence_data.dat};
  \addlegendentry{Statevector}
  \addplot [thick, color=CBred, mark=x, mark size=3pt, mark options={solid}]
    table [x=step, y=shots_err] {ibm_data/convergence_data.dat};
  \addlegendentry{Ideal shots}
  \addplot [thick, dashed, color=CBgreen, mark=triangle, mark size=3pt, mark options={solid}]
    table [x=step, y=ibm_err] {ibm_data/convergence_data.dat};
  \addlegendentry{IBM QPU}
\end{axis}
\end{tikzpicture}
    \caption{Full state convergence of the nonlinear airfoil simulation to a steady state computed as the mean squared error (MSE) for three different scenarios: statevector simulation, ideal simulation with a fixed number of shots and result recovery with tomography, and execution on the \code{ibm\_boston} quantum computer.}
    \label{fig:qlbm_convergence}
\end{figure}
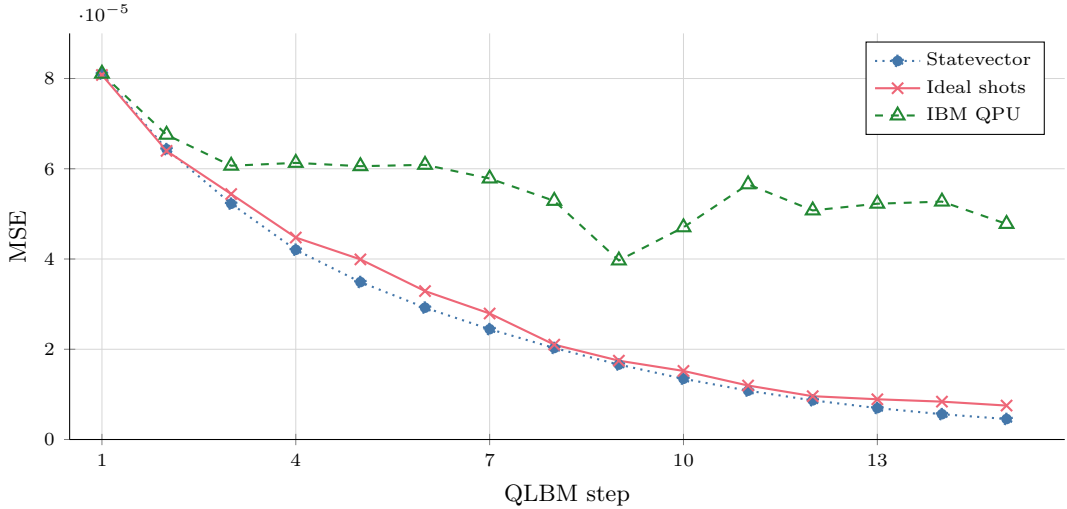

In Figure~\ref{fig:qlbm_convergence} we further compare the convergence of all three fields to the steady state of the model, obtained by executing the algorithm on a state vector simulator for a sufficiently long time. The state-vector simulation is shown in blue in Fig.~\ref{fig:qlbm_convergence}, an ideal simulation with shot noise included is shown in red, and the results of the circuit execution on the noisy IBM QPU is shown in green. While the results are affected by noise, the algorithm nevertheless continues to convergence against mean squared error, reducing the error with subsequent steps.
\section{A look into the future}%
\label{sec:future}

We have presented a comprehensive model for designing a flexible quantum algorithm for the lattice Boltzmann method. The core model -- collision, propagation, and integration -- is efficient in terms of gate complexity and relatively simple to set up. For a subset of linear physics problems amenable to the LBM treatment, there is even a good case to be made for a fully unitary, end-to-end complete quantum algorithm.

Yet some pain points remain. The two major obstacles on the road to industrial-scale quantum LBM simulations are the non-unitarity of some of the operators in the general case, and the difficulties in modelling nonlinear physics. 

Non-unitarity leads to the dissipation of amplitudes within the time-stepping scheme, thus diminishing the probability of measuring meaningful data at the end of a simulation run. The options here are varied: For one, it may be possible to unitarize at least some of the operators. This can be done directly or by using quantum machine learning with a unitary ansatz \cite{lacatus2025surrogatequantumcircuitdesign,Itani2025QMLLBM}. Recent research indicates that this is possible for linear cases, where the equilibrium distribution functions of the LBM can be approximated by neglecting the nonlinear terms $\bigO(u^2)$. This is especially true at low Reynolds numbers and low velocities. The quantum circuits produced using this technique are shallower, unitary, and have low error rates. Despite this success, the generality of the learned operator and its efficiency for simulations involving several time steps are yet to be shown. The second option is to boost the data with amplitude amplification. Recent work \cite{Zecchi2025AmplitudeAmplification} focused on the application of oblivious amplitude amplification to non-unitary operators using block-encoding or LCU, where the initial statevector is not known. This work shows that while the application of amplification introduces errors in the relative amplitudes of the statevector, an approximate version may result in viable solutions. Lastly, it may be possible to devise alternative methods for data encoding rather than pure amplitudes. These methods are not mutually exclusive, and we may end up using all of them in some form.

Nonlinearity too is tied to the amplitude encoding in that the available operators for direct amplitude manipulation are necessary unitary, and thus linear. However, a sizable body of literature is emerging in how to tackle nonlinear operators and/or how to change the encoding to allow more flexibility beyond unitary operators.

\FloatBarrier

\printbibliography

\appendix
\section{Hydrodynamics and acoustics -related lattice Boltzmann models}%
\label{sec:appendix-lbm}

\subsection{LBM for hydrodynamic and acoustic models}

In the case of Navier-Stokes hydrodynamics, the equilibrium part is given by a quadratic expansion of the exact Maxwellian, leading to~\cite[Ch.~4]{kruger2016lattice}
\begin{equation}
    f_{\alpha}^{eq}(x,t) = \rho W_{\alpha}\left( 1 + \frac{\bm{c}_{\alpha} \cdot \bm{u}}{c_s^2}+\frac{(\bm{c}_{\alpha} \cdot \bm{u})^2}{2c_s^4}-\frac{\bm{u}\cdot\bm{u}}{2c_s^2}\right),
    \label{eq:navier-stokes}
\end{equation}
while the non-equilibrium part is defined as $f_{\alpha}^{neq}=f_{\alpha}-f_{\alpha}^{eq}$. 
The macroscopic quantities, i.e.  the density $\rho$ and the momentum $\rho \bm{u}$ are computed as follows:
\begin{equation}
    \sum_i f_i ^{eq} = \rho, \quad \sum_i  \bm{c}_i f_i ^{eq} = \rho \bm{u}
\end{equation}
This expression holds for a weakly compressible athermal fluid flows. It is able to capture both shear and acoustic disturbances.
The corresponding macroscopic hydrodynamic governing equations are
\begin{eqnarray}
\frac{\partial \rho}{\partial t} + \nabla \cdot ( \rho \bm{u}) & = & 0 \\
\frac{\partial \rho \bm{u}}{\partial t} + \nabla \cdot ( \rho \bm{u} \bm{u} + p \bm{I}) & = & \nu \nabla^2 \bm{u}
\end{eqnarray}
where the pressure $p$ is given by the LBM-native equation of state $p = \rho c_s^2$, where $c_s$ is the characteristic lattice speed.
The physical kinetic viscosity $\nu$ of the fluid is related to the lattice speed $c_s$, the relaxation parameter $\tau$, the mesh size $\Delta x$ and the time step $\Delta t$ via the following relation
\begin{equation}
    \nu = \frac{( \tau -1/2)}{ c_s^2}\frac{\Delta x^2}{\Delta t}
\end{equation}

This model can be modified to handle (nearly) incompressible flows~\cite{he1997lattice}, in which density is split as $\rho = \rho_0 + \rho'$, where $\rho_0$ and $\rho' \ll \rho_0$ denote the base flow density and the density fluctuations, respectively.
Introducing this splitting into Eq.~\eqref{eq:navier-stokes}, one obtains
\begin{equation}
    f_{\alpha}^{eq}(x,t) =  W_{\alpha}\left( \rho ' + \rho_0 \left[ \frac{\bm{c}_{\alpha} \cdot \bm{u}}{c_s^2}+\frac{(\bm{c}_{\alpha} \cdot \bm{u})^2}{2c_s^4}-\frac{\bm{u}\cdot\bm{u}}{2c_s^2} \right] \right),
    \label{eq:navier-stokes-incompressible}
\end{equation}
with the following macroscopic unknowns
\begin{equation}
    \sum_i f_i ^{eq} = \rho', \quad \sum_i  \bm{c}_i f_i ^{eq} = \rho_0 \bm{u}
\end{equation}
Here, it is assumed that terms  scaling as $\rho' \bm{u}$ are negligible.
In this incompressible approximation, density fluctuations are decoupled from velocity. The equilibrium functions are linear in $\rho '$ but quadratic in $\mathbf{u}$.

The associated macroscopic equations are 
\begin{eqnarray}
\frac{1}{c_s^2}\frac{\partial P}{\partial t} + \nabla \cdot  \bm{u} & = & 0 \\
\frac{\partial\bm{u}}{\partial t} + \bm{u} \cdot \nabla  \bm{u}  & = & - \nabla P + \nu \nabla^2 \bm{u} + \nu \nabla ( \nabla \cdot \bm{u})
\end{eqnarray}
with $P= p/\rho_0$. This system is proven to be similar to the artificial compressibility method, and approximate to the incompressible Navier-Stokes equation~\cite{he1997lattice,he2002comparison}:
\begin{eqnarray}
 \nabla \cdot  \bm{u} & = & 0 + O(M^2) \\
\frac{\partial\bm{u}}{\partial t} + \bm{u} \cdot \nabla  \bm{u}  & = & - \nabla P + \nu \nabla^2 \bm{u} + O(M^3) ,
\end{eqnarray}
where the error appears in terms of the Mach number $M = u/c_s$.
It is worth noting that for the quantum algorithm for the OSSLBM, this expression for the equilibrium is much better suited than the one proposed in \cite{guo2000lattice}.

The equilibrium function can be further linearized considering acoustic disturbances, for which the velocity can be split as $\bm{u} = \bm{u}_0 + \bm{u}'$, with $\bm{u}_0$ and $\bm{u}'$ the base flow velocity and the acoustic disturbance velocity field, respectively. According to the linear acoustic assumptions, one has $\rho' \ll \rho_0$.
For the sake of simplicity, the momentum can be split as $ \rho \bm{u} = (\rho \bm{u})_0 + (\rho \bm{u})'$, where $(\rho \bm{u})' = \rho_0 \bm{u}' + \rho ' \bm{u}_0 + \rho ' \bm{u}' \simeq \rho_0 \bm{u}' + \rho ' \bm{u}_0$, yielding
\begin{equation}
    \sum_i f_i ^{eq} = \rho', \quad \sum_i  \bm{c}_i f_i ^{eq} = (\rho \bm{u})'
\end{equation}
Neglecting all nonlinear terms to recover the classical linear acoustics equations, one has
\begin{eqnarray}
    f_{\alpha}^{eq}(x,t) & =&  W_{\alpha} \left[\rho' \left( 1 + \frac{\bm{c}_{\alpha} \cdot \bm{u}_0}{c_s^2}+\frac{(\bm{c}_{\alpha} \cdot \bm{u}_0)^2}{2c_s^4}-\frac{\bm{u}_0\cdot\bm{u}_0}{2c_s^2}\right) \right.\nonumber \\
    & & + \left.  \rho_0 \left(  \frac{\bm{c}_{\alpha} \cdot \bm{u}'}{c_s^2}+\frac{(\bm{c}_{\alpha} \cdot \bm{u}_0)(\bm{c}_{\alpha} \cdot \bm{u}')}{c_s^4}-\frac{\bm{u}_0\cdot\bm{u}'}{c_s^2}  \right) \right],
    \label{eq:acoustics}
\end{eqnarray}
The associated macroscopic equation are
\begin{eqnarray}
\frac{\partial \rho'}{\partial t} + \nabla \cdot ( \rho \bm{u})' & = & 0 \\
\frac{\partial (\rho \bm{u})'}{\partial t} + \nabla \cdot ( (\rho \bm{u})_0 \bm{u}'+ (\rho \bm{u})' \bm{u}_0 + p' \bm{I}) & = & \nu \nabla^2 \bm{u}'
\end{eqnarray}

\subsection{LBM for shallow water -based physical models}

The classical shallow water equations are obtained by averaging the Navier-Stokes equations in the water bed depth direction.
Introducing the water  height $h$ and the depth-averaged fluid velocity 2D vector $\bm{u}$, the equilibrium functions are defined as (see e.g. \cite{zhou2002lattice,geveler2010lattice,de2017central,maquignon2022simplified,de2023comparison}):
\begin{equation}
 \displaystyle   f_i^{eq} = \begin{cases}  
    \displaystyle h - \frac{5 g h^2}{2 c_s^2} - \frac{2 h}{c_s^2} (\bm{u} \cdot \bm{u}) & i=0 \\
    \displaystyle \frac{ g h^2}{2 c_s^2} + \frac{h}{c_s^2}(\bm{c}_i \cdot \bm{u}) + \frac{9 h}{2 c_s^4}(\bm{c}_i \cdot \bm{u})^2 - \frac{ h}{2 c_s^2} (\bm{u} \cdot \bm{u}) & i= 1,4 \\
   \displaystyle  \frac{1}{4} \left[ \frac{ g h^2}{2 c_s^2} + \frac{h}{c_s^2}(\bm{c}_i \cdot \bm{u}) + \frac{9 h}{2 c_s^4}(\bm{c}_i \cdot \bm{u})^2 - \frac{ h}{2 c_s^2} (\bm{u} \cdot \bm{u}) \right] & i= 5,8
       \end{cases}
\end{equation}
where $g$ is the gravitational acceleration.
The reconstructed macroscopic quantities are the water bed height and the mass flux
\begin{equation}
    \sum_i f_i ^{eq} = h, \quad \sum_i  \bm{c}_i f_i ^{eq} = h \bm{u}
\end{equation}
The associated macroscopic equation for the flow of flat bed are
\begin{eqnarray}
\frac{\partial h}{\partial t} + \nabla \cdot ( h \bm{u}) & = & 0 \\
\frac{\partial (h \bm{u})}{\partial t} + \nabla \cdot ( h \bm{u} \bm{u}) & = & - g \nabla \left( \frac{h^2}{2} \right) +  \nu \nabla \cdot ( h \nabla \bm{u})
\end{eqnarray}
An additional forcing term can be added to account for variations in the bed elevation.

This system can be linearized to investigate weak surface wave dynamics.
Introducing the following fluctuating variables:
\begin{equation}
    \sum_i f_i ^{eq} = h', \quad \sum_i  \bm{c}_i f_i ^{eq} = (h \bm{u})' = h \bm{u} - h_0 \bm{u}_0 \sim h_0 \bm{u}' + h' \bm{u}_0
\end{equation}
where $h_0$ and $\bm{u}_0$ are the base flow quantities, one has
\begin{equation}
 \displaystyle   f_i^{eq} = \begin{cases}  
    \displaystyle  \left( 1 - \frac{5 g h_0}{c_s^2} + \frac{2}{c_s^2} (\bm{u}_0 \cdot \bm{u}_0) \right) h' - \frac{4}{c_s^2} \bm{u}_0 \cdot (h \bm{u})' & i=0 \\
    \displaystyle  \left( \frac{ g h_0}{c_s^2} -\frac{9 }{2 c_s^4}(\bm{c}_i \cdot \bm{u}_0)^2 - \frac{ 1}{2 c_s^2} (\bm{u}_0 \cdot \bm{u}_0) \right) h' & i= 1,4 \\
   \displaystyle  + \left( \frac{\bm{c}_i+\bm{u}_0}{c_s^2} + \frac{9 (\bm{c}_i \cdot \bm{u}_0)}{c_s^4} \bm{c}_i  \right) \cdot (h \bm{u})' \\
   \displaystyle \left( \frac{ g h_0}{c_s^2} -\frac{9 }{2 c_s^4}(\bm{c}_i \cdot \bm{u}_0)^2 - \frac{ 1}{2 c_s^2} (\bm{u}_0 \cdot \bm{u}_0) \right) \frac{h'}{4} & i= 5,8 \\
   \displaystyle  + \frac{1}{4}\left( \frac{\bm{c}_i+\bm{u}_0}{c_s^2} + \frac{9 (\bm{c}_i \cdot \bm{u}_0)}{c_s^4} \bm{c}_i  \right) \cdot (h \bm{u})' \\
\end{cases}
\end{equation}
The corresponding set of macroscopic equations is
\begin{eqnarray}
\frac{\partial h'}{\partial t} + \nabla \cdot ( h \bm{u})' & = & 0 \\
\frac{\partial (h \bm{u})'}{\partial t} + \nabla \cdot ( (h \bm{u})' \bm{u}_0 + (h \bm{u})_0 \bm{u}') & = & - g \nabla  h' +  \nu \nabla \cdot ( h' \nabla \bm{u}_0 + h_0 \nabla \bm{u}') .
\end{eqnarray}

\section{Function tomography of quantum states}%
\label{sec:appendix-tomography}

Here we introduce and describe a way to perform quantum state tomography, which is efficient, provided certain sparsity assumptions on the function encoded in the quantum amplitudes are satisfied.

Let $\vert \psi\rangle$ be a $n$-qubit wavefunction prepared on a quantum computer. We write it as:
\begin{equation}
    \vert\psi\rangle = \sum_{i=0}^{2^n-1} f(x_i) \vert i\rangle,
\end{equation}
where $\vert i\rangle$ is the $i$-th computational basis state, and $x_i$ is the $i$-th grid point in the discretized domain of some function $f(x)$. For example, if $f$ is defined on the $[0,1]$ interval, we can put $x_i = i / 2^n$. Here, for simplicity, we assume that $f$ is a 1D function, but the approach trivially generalizes to higher dimensions.

We now assume that $f(x)$ can be expressed as the following finite sum
\begin{equation}
    f(x) = \sum_{j=1}^m a_j g_j(x),
\end{equation}
where $g_j(x)$ is some predetermined (not necessarily orthgonal) function basis of \emph{real} functions and $a_j$ are also real. Crucially, we further assume sparsity, that is, $m \ll 2^n$. No further assumptions on the functions $g_j(x)$ are imposed. For the quantum state to be properly normalized, however, the vector of amplitudes $f(x)$ must have an unit norm. This condition can be written as follows:
\begin{equation}
    W(a) = \sum_j \sum_k a_j G_{jk} a_k = 1,
\end{equation}
where 
$G_{j,k}$ is the inner product of $g_j$ and $g_k$ on the discretized domain. The choice of orthonormal functions $g_j$ simplifies the expression to $W(a) = \sum_{j=1}^m a_j^2$.

The goal of function tomography is to efficiently find the coefficients $a_j$ having only the access to the projective measurements from the quantum state $\vert \psi \rangle$.

\subsection{Recovering the absolute value of the encoded function}

Let us first consider a simpler problem of performing function tomography to extract only the absolute values  $|f(x)|$ of the encoded function. It can be used when it is known that $f(x) \geq 0$, or that the sign can be recovered subsequently in a different way.

Let $P(i)$ be the probability distribution of quantum states obtained from $\vert\psi\rangle$ by measuring in the computational ($\z$) basis. $P(i)$ is equal to number of times the state $\vert i\rangle$ was observed divided by the number of performed shots. Let $a$ be the vector of (current values of) the variational coefficients  $a_j$ and let us denote the linear combination of the functions $g_j(x)$ it specifies as $f_a(x)$. The goal is to fix $a$ such that $f_a(x) \approx f(x)$. Toward this goal, following \cite{Torlai2023}, we can compute the distance between $P(i)$ and the probability of state $|i\rangle$ sampled with $f_a(x)$ in terms of the Kullback-Leibler divergence between these two distributions, or, equally, the negative log-likelihood:
\begin{equation}\label{eq:tomography_kl_loss}
    L(a) = - \sum_i P(i) \log\left( \frac{|f_a(x_i)|^2}{W(a) P(i)}\right).
\end{equation}
This loss function can be efficiently minimized with a gradient based approach:
\begin{equation}
    \frac{\partial L}{\partial a_j} = -2 \sum_{i} P(i) g_j(x_i) / f_a(x_i) + 2\sum_{k=1}^m a_k G_{j, k}  / W(a), 
\end{equation}
where the optimization is to be performed under the constraint of $W(a)=1$, to ensure the proper quantum norm of the solution.

Note that the sum over computational states $i$ in \eqref{eq:tomography_kl_loss} goes only over the states which were measured, and therefore the loss function has no more terms than the amount of shots taken. This ensures that the cost function can always be computed, even in systems with large number of qubits $n$.

\subsection{Recovering the complete real-valued encoded function} 

To perform tomography which reconstructs the complete real function $f(x)$, including the relative signs in the amplitude, we combine the above approach (itself using a similar cost function to matrix product state (MPS) tomography \cite{Torlai2023,Termanova_2024}) with properties of the real state tomography \cite{li2025efficientcircuitbasedquantumstate,song2025hadamardrandomforestreconstructing,Song_2025}.

To recover a complete state, in all tomography approaches measurements in multiple bases must be performed. Real-state tomography approaches \cite{li2025efficientcircuitbasedquantumstate,song2025hadamardrandomforestreconstructing,Song_2025} are based on the observation that if the function encoded in the amplitudes is known to be real-valued, it then suffices to measure using a number of bases scaling only linearly in the number of qubits $n$ (as opposed to exponential number in the general case). Concretely, in addition to the $\z$-basis measurement it suffices to measure in $n$ additional bases only, obtained by rotating on of the qubits to the local $\x$ basis.
Note that there is no possibility to recover the overall sign, as the functions $f(x)$ and $-f(x)$ differ by a global phase only.

Let now $P_k^{(X)}(i)$ be the probability distribution, measured from a quantum state $\vert\psi\rangle$, where $k$-th qubit (counting from $0$) was in $\x$ basis.
Then adapting the above real-state tomography approach to our task of function tomography, that is, determining coefficients in a fixed finite basis of functions $g_j$, we introduce $n$ additional terms in the cost function ($0\leq k<n)$:
\begin{equation}
    L_k^{(X)}(a) = - \sum_i P_k^{(X)}(i) \log\left( \frac{|f_a(x_i) \pm f_a(x_{i + 2^k \mod 2^n})|^2}{2W(a) P_k^{(X)}(i)}\right),
\end{equation}
where the plus or minus sign is determined by whether the $k$-th measurement is $0$ or $1$ respectively.

We note that function tomography in the sense described above (fixing coefficients in a function basis) exists for some special sets of basis functions, albeit with a crucial technical and practical difference. To wit, when the basis $g_j$ is assumed to to be Fourier \cite{huang2025realfourierspacereadout}, or Chebyshev polynomials \cite{su2025efficientquantumstatetomography} methods of extracting the coefficients $a_j$ have been described. In contrast to our approach, however, these are measured directly on the quantum device and require the addition of a complex quantum circuit which acts on the state $\vert\psi\rangle$ before measurements. This renders the circuit too deep in practice for NISQ devices. On the other hand, however, ideal simulator results with the above methods have demonstrated that such function tomography can in principle represent sufficiently accurately various complicated fluid flows and its regimes, despite the limitation of representing the result as a finite sum of functions.

Finally, to account to possible discontinuities in the function due to the presence of the object, when performing the tomography over a flow, additional grid transformations can be performed. This may allow to use fewer basis functions in the tomography, and therefore make the variational fitting simpler. In particular, for the experiment in Section~\ref{sec:example_airfoil} a rectangular object of size $2a\times 2b$ was used, whose center was located in the origin of the coordinate system. Using a mapping $(x,y) \mapsto (x,y) \cdot \max(r-1,0)/r$, where $r=\max\{|x/a|, |y/b|\}$, the coordinate system grid around the object was transformed and the object itself mapped to a single point (see also \cite{liseikin1999grid}). It was numerically confirmed that this simple transformation allowed to reduce the number of basis functions needed for the tomography. For more complicated objects and flows conformal maps can be used instead \cite{FLORYAN1986221}. 

\end{document}